\documentclass[journal]{IEEEtran}
\ifCLASSINFOpdf
\else
\fi
\usepackage{amssymb}
\usepackage{gensymb}
\usepackage{graphicx}
\usepackage{amsmath}
\usepackage{float}
\usepackage{algorithm}
\usepackage{algpseudocode}

\usepackage{makecell}
\usepackage{cite}

\usepackage{xcolor}
\usepackage{threeparttable}

\begin{document}
%
\title{Rate-Splitting Multiple Access for Multigroup Multicast 
and Multibeam Satellite Systems}
%
%
%

\author{Longfei~Yin,
        Bruno~Clerckx,~\IEEEmembership{Senior Member,~IEEE}
        
\thanks{
L. Yin and B. Clerckx are with the Communications and Signal Processing Group, Department of Electrical and Electronic Engineering, Imperial College London, London, SW7 2AZ, U.K (e-mail: longfei.yin17@imperial.ac.uk; b.clerckx@imperial.ac.uk).
This work has been partially supported by the U.K. Engineering and Physical Sciences Research Council (EPSRC) under grant EP/R511547/1. This paper has been partially submitted for conference publication \cite{9145200}.}

}

\maketitle
\begin{abstract}
This work focuses on the promising Rate-Splitting Multiple Access (RSMA) and its beamforming design problem to achieve max-min fairness (MMF) among multiple co-channel multicast groups with imperfect channel state information at the transmitter (CSIT). 
Contrary to the conventional linear precoding (NoRS) that relies on
fully treating any residual interference as noise, we consider a novel multigroup multicast beamforming strategy based on RSMA. RSMA relies on linearly precoded Rate-Splitting (RS) at the transmitter and Successive Interference Cancellation (SIC) at the receivers, and has recently been shown to enable a flexible framework for non-orthogonal transmission and robust interference management in multi-antenna wireless networks. 
In this work, we characterize the MMF Degrees-of-Freedom (DoF) achieved by RS and NoRS in multigroup multicast with imperfect CSIT and demonstrate the benefits of RS strategies for both underloaded and overloaded scenarios.
Motivated by the DoF analysis, we then formulate a generic transmit power constrained optimization problem to achieve MMF rate performance. 
The superiority of RS-based multigroup multicast beamforming compared with NoRS is demonstrated via simulations 
in both terrestrial and multibeam satellite systems. 
In particular, due to the
characteristics and challenges of multibeam satellite communications, our proposed RS strategy is shown promising 
to manage its interbeam interference. 
\end{abstract}
%


%
\IEEEpeerreviewmaketitle

\section{Introduction}
%
%
%
%

 

\IEEEPARstart{W}{ith} the proliferation of mobile data and multimedia traffic, demands for massive connectivity and content-centric services are continuously rising. Examples include audio/video streaming, advertisements, large scale system updates, localized services and downloads, etc.
Spurred by such requirements, wireless multicasting has attracted a widespread research attention.
It is a promising solution to deliver the same message to a group of recipients.
In a more general scenario, which is known as multigroup multicasting, distinct contents are simultaneously transmitted to multiple co-channel multicast groups.  
Since the available spectrum is aggressively reused towards spectrum efficient and high throughput wireless communications, 
interference mitigation techniques are of particular importance.
\subsection{Related Works}
Multicast beamforming is firstly considered in \cite{sidiropoulos2006transmit} with single-group setup.
Then, the problem is extended to multigroup multicasting in \cite{karipidis2008quality} where the beamforming design is investigated in two optimization perspectives, namely the QoS constrained total transmit power minimization (QoS problem) and the power constrained max min fairness (MMF problem). Both formulations are shown to be NP-hard, containing the multiuser unicast and the single-group multicast as extreme cases. The combination of Semi-Definite Relaxation (SDR) and Gaussian randomization, together with bisection search algorithm are elaborated to generate feasible approximate solutions. 
Alternatively, a convex-concave procedure (CCP) \cite{lanckriet2009convergence} algorithm is demonstrated to provide better performance. However, its complexity increases dramatically as the problem size grows. 
In \cite{chen2017admm}, a low-complexity algorithm for multigroup multicast beamforming design based on alternating direction method of multipliers (ADMM) together with CCP is proposed for large-scale wireless systems. 
Moreover, the multigroup multicast beamforming is extended to many other scenarios, including the per-antenna power constraint addressed in \cite{christopoulos2014weighted},  Cloud-RANs with wireless backhaul \cite{hu2016multicast}, coordinated beamforming in multi-cell networks \cite{xiang2012coordinated}, cache aided networks \cite{tao2016content} and massive MIMO \cite{sadeghi2017max}.

One practical application of multigroup multicast is found in multibeam satellite communication systems.
In recent years, the multibeam satellite system has received considerable research attentions due to its ubiquitous coverage, and full frequency reuse across multiple narrow spot
beams towards high throughput. 
Based on state of the art technologies in DVB-S2X \cite{DVB}, each spot beam of the satellite serves more than one user simultaneously by transmitting a single coded frame. 
Since different beams illuminate different group of users \cite{wang2019multicast}, this promising SatCom system follows the aforementioned physical layer
(PHY) multigroup multicast transmission. 
In the literature of multibeam satellites, a generic iterative algorithm is proposed in \cite{zheng2012generic} 
to design the precoding and power allocation alternatively in a time division multiplexed (TDM) scheme considering single user per beam.
Then, multigroup multicast is considered.
\cite{christopoulos2015multicast} proposes a frame-based precoding problem for multibeam multicast satellites.
Optimization of system sum rate is considered under individual power constraints via an alternating projection technique with a semidefinite relaxation (SDR) procedure, which is adequate for small to medium-coverage areas. 
In \cite{joroughi2016generalized},  a two-stage low complex precoding
design for multibeam multicast satellite systems is proposed.
The first stage minimizes the inter-beam interference, while the second stage enhances intra-beam SINR. 
\cite{wang2019multicast} studies the sum rate maximization problem in multigateway multibeam satellite systems considering feeder link interference.
Leakage-based MMSE and SCA-ADMM algorithm are used to compute precoding vectors locally with limited coordination.

It is noted that all aforementioned works rely on the conventional multigroup multicast linear precoding. Each user decodes its desired stream while treating all the other interference streams as noise. 
The advantage of this conventional scheme lies in exploiting the spatial degrees of freedom provided by multiple antennas using low complexity transmitter-receiver architecture.
However, its effectiveness severely depends on the network load and the quality of channel state information at the transmitter (CSIT).
As precoders are designed based on the channel knowledge,
CSIT inaccuracy can result in an inter-group interference problem which is detrimental to the system performance.
Another limitation is that the conventional linear precoding is able to eliminate inter-group interference only when the number of transmit antennas is sufficient. Otherwise, it fails to do so in overloaded systems \cite{joudeh2017rate}. 
For example, rate saturation occurs in overloaded systems. 

In this work, we depart from this conventional scheme (denoted as NoRS in the sequel) and introduce the powerful Rate-Splitting Multiple Access (RSMA) to the multigroup multicast setup. RSMA relies on one-layer or multiple-layer linearly
precoded Rate-Splitting (RS) at the transmitter and Successive Interference Cancellation (SIC) at the receivers \cite{clerckx2016rate, mao2018rate}. It contains NoRS and Non-Orthogonal Multiple Access (NOMA) as a special case \cite{clerckx2019rate,mao2018rate}. 
The key of RS-based multigroup multicast beamforming is to divide each group-intended message into a common part and a private part. The common parts are jointly encoded into a common stream, while the private parts are separately encoded into private streams.
At receiver sides, the common stream is firstly decoded by all the users and then removed through SIC. Next, each user decodes its desired private stream and treats the remaining interference as noise. This characteristic enables RS to partially decode the interference and partially treat the interference as noise \cite{clerckx2016rate}. 
The benefits of RS have been investigated in a wide range of multi-antenna setups, namely multiuser unicast transmission with perfect CSIT \cite{mao2018rate,ahmad2019interference,clerckx2019rate,zhang2019cooperative }, imperfect CSIT \cite{joudeh2016sum, hao2015rate, joudeh2016robust, piovano2017optimal, dai2016rate,mao2019beyond,caus2018exploratory}, multigroup multicast transmission \cite{joudeh2017rate, chen2019joint,yalcin2019rate, tervo2018multigroup}, as well as
superimposed unicast and multicast transmission \cite{mao2019rate}, etc.
According to the analysis and simulations, \cite{mao2018rate} shows that RS is more robust to the influencing factors such as channel disparity, channel orthogonality, network load, and quality of CSIT. 
For imperfect CSIT, the sum Degrees-of-Freedom (DoF) and MMF-DoF of underloaded MU-MISO system are studied in \cite{joudeh2016sum} and \cite{joudeh2016robust}. 
Compared with NoRS, RS is demonstrated to further exploit spatial dimensions. 
Of particular interest to this work is \cite{joudeh2017rate}, the employment of RS in multigroup multicast beamforming was first proposed. An MMF problem based on RS is formulated and solved by the Weighted Minimum-Mean Square Error (WMMSE) approach \cite{christensen2008weighted}. The superiority of RS
under perfect CSIT is shown in overloaded multigroup multicast systems.
\subsection{Contributions}
 In this work, motivated by further exploring the benefits of RS for multigroup multicast beamforming, we consider both underloaded and overloaded regimes with imperfect CSIT.
Furthermore, its application to multibeam satellite systems is investigated. The main contributions are as follows:
 \begin{itemize}
 \item 
First,
this paper is a follow-up and extension of \cite{joudeh2017rate}, which
 studied RS for multigroup multicast under perfect CSIT and overloaded systems.
This is the \textit{first paper} to study RS for multigroup multicast in the more general
and practical scenario of imperfect CSIT in both underloaded and overloaded systems.

\item
Second,
the MMF-DoF
 of RS and NoRS in multigroup multicast with imperfect CSIT is characterized.
The MMF-DoF, also known as max-min fair multiplexing gain, corresponds to the maximum multiplexing gain that can be simultaneously achieved by all multicast groups.
It reflects the pre-log factor of MMF-rate at high SNR.
This is the \textit{first work} on DoF analysis for multigroup multicast in the presense of imperfect CSIT.
In \cite{joudeh2017rate}, MMF-DoF gains of RS with perfect CSIT were only observed
in overloaded systems. 
In this work with imperfect CSIT setting, RS is shown to provide
MMF-DoF gains in both underloaded and overloaded systems.
Through residual interference and group partitioning analysis,
RS is a shown to be 
 more flexible than NoRS to overcome the residual interference caused by imperfect CSIT.
 By adjusting the common stream and private streams, we can determine how much interference to be decoded and how much to be treated as noise.
Due to the existence of common part, RS provides extra gains and avoids the saturating performance at high SNR.

 \item 
Third, motivated by the benefits of RS over NoRS from a DoF perspective, a MMF beamforming optimization problem is then formulated to see whether the DoF gain translates into
rate gain.
To this end, the design of RS for MMF rate maximization at finite SNR is further investigated.
  This is the \textit{first work} on the optimization of RS-based multigroup multicast with imperfect CSIT.
Solving the MMF problem with imperfect CSI via SAA and WMMSE is for the \textit{first time} studied. 
Optimum MMF Ergodic Rate can be obtained by optimizing the defined short-term MMF Average Rate (AR) over a long sequence of channel estimates. 
 The formulated problem is general enough to cope with flexible power constraints, namely a total power constraint (TPC) and per-antenna power constraints (PAC).
 Through simulation results, the DoF benefits of RS over NoRS translate into rate benefits at finite SNR and RS is shown to outperform NoRS in a wide range of setups.
All the simulation results are inline with the derived theoretical MMF-DoFs results.
 Considering imperfect CSIT, we show that RS for multigroup multicast brings spectral efficiency gains over NoRS in both underloaded and overloaded scenarios. 
 This contrasts with the perfect CSIT setting of \cite{joudeh2017rate}, where RS was shown to provide significant spectral efficiency benefits in the overloaded scenarios only.
\item
Fourth, the proposed RS framework is applied
to a multibeam satellite setup and results confirm the significant performance gains over traditional techniques. 
Since multibeam satellite communication systems aim to operate with full frequency reuse to enable higher throughput, interference management techniques should be employed. 
Based on state of the art technologies in DVB-S2X,
each spot beam of the satellite serves more than one user simultaneously by transmitting a single coded frame. 
So, this multibeam multicasting follows the physical layer (PHY) of a multigroup multicast transmission. 
Different from \cite{8491094}, which studies RS in a two-beam satellite system adopting TDM scheme in each beam, and
\cite{8385504} which focuses on the sum-rate optimization and low complexity RS precoding design with perfect CSIT,
we consider a novel RS-based multibeam
multicast beamforming in this paper
and formulate a per-feed power constrained MMF problem with different CSIT qualities.
RS framework is shown
very promising for multibeam satellite systems to manage its inter-beam interference, taking into account practical challenges such as CSIT uncertainty, per-feed constraints, hot spots, uneven user distribution per beam, and overloaded regimes.
Simulation results confirm the significant performance gains over traditional techniques.

\end{itemize}

\subsection{Organization and Notations}
The rest of this paper is organized as follows. The system model, CSIT assumptions and RS for multigroup multicast are introduced in Section II. 
Section III investigates the MMF-DoF of both NoRS and RS.
Section IV
provides details of the problem formulation. A modified WMMSE approach is designed to optimize the MMF RS-based multigroup multicast beamforming. 
Applications to both terrestrial and multibeam satellite systems are given in Section V via simulations. Finally, Section VI concludes this work.

Notations: In the remainder of this paper, boldface uppercase, boldface lowercase and standard letters denote matrices, column vectors, and scalars respectively. 
$\mathbb{R}$ and $\mathbb{C}$ denote the real and complex domains. 
The real part of a complex number $x$ is given by $\mathcal{R}\left (x  \right )$. 
$\mathbb{E}\left ( \cdot  \right )$ is the expectation of a random variable. 
The operators $\left ( \cdot  \right )^{T}$ and $\left ( \cdot  \right )^{H}$ denote the transpose and the Hermitian transpose. $\left | \cdot  \right |$ and $\left \| \cdot  \right \|$ denote the absolute value and Euclidean norm respectively.

\section{System Model}
We consider a multigroup multicasting downlink MISO system. The transmitter is equipped with $N_{t}$ antennas, serving $K$ single-antenna users which are grouped into $M \left ( 1\leq M\leq K \right )$ multicast groups. 
The users within each group desire the same multicast message.
The messages are independent amongst
different groups. 
Let $\mathcal{G}_{m}$ denote the set of users belonging to the $m$-th group, for all $ m\in \mathcal{M} = \left \{ 1\cdots M \right \}$. 
The size of group-$m$ is $G_{m}=\left | \mathcal{G}_{m} \right |$.
We assume that each user belongs to only one group, thus $\mathcal{G}_{i} \ \cap \ \mathcal{G}_{j}  =\emptyset$, for all $ i,j\in \mathcal{M}$, $ i\neq j$.
Let $\mathcal{K} = \left \{ 1\cdots K \right \}$ denote the set of all users, i.e. $\cup _{m \in \mathcal{M}} \ \mathcal{G}_{m} = \mathcal{K}$.
In this model, 
the signal received at user-$k$ writes as $y_{k} = \mathbf{h}_{k}^{H}\mathbf{x} + n_{k}$,
 $\forall k\in \mathcal{K}$, where
 $\mathbf{x}\in \mathbb{C}^{N_{t}\times 1}$ is the transmitted signal,
 $\mathbf{h}_{k}\in \mathbb{C}^{N_{t} \times 1}$ is a channel vector between the transmitter and the $k$-th user. $\mathbf{H} \triangleq \left [ \mathbf{h}_{1}, \cdots,\mathbf{h}_{K} \right ]$ is the composite channel. 
 $n_{k}\sim \mathcal{CN}\big ( 0,\sigma _{n,k}^{2} \big )$ represents the Additive White Gaussian Noise (AWGN) at user-$k$, which is independent and identically distributed (i.i.d) across users with zero mean and variance $\sigma _{n,k}^{2}$. Without loss of generality, unit noise variances are assumed, i.e. $\sigma _{n,k}^{2} = \sigma _{n}^{2}=1$.

\subsection{Transmission Schemes}
The RS strategy for multigroup multicasting is described as follows. There are overall $M$ messages $W_{1},\cdots,W_{M}$ intended to users in $\mathcal{G}_{1},\mathcal{G}_{2},\cdots ,\mathcal{G}_{M}$ respectively. Each message is split\footnote{The readers are referred to \cite{clerckx2019rate, clerckx2016rate, mao2018rate, joudeh2016sum, hao2015rate, joudeh2016robust} for a general introduction to multi-antenna rate-splitting.} 
into a common part and a private part, i.e. $W_{m}\rightarrow \left \{ W_{m,c} , W_{m,p} \right \}$. All the common parts are packed together and encoded into a common stream shared by all groups, i.e. $\left \{ W_{1,c} \cdots W_{M,c}\right \}\rightarrow s_{c}$, while the private parts are encoded into private streams for each group independently, i.e. $W_{m,p}\rightarrow s_{m}$.
As a consequence, 
the vector of symbol streams to be transmitted is $\mathbf{s}=\left [ s_{c} ,s_{1},\cdots ,s_{M}\right ]^{T}\in \mathbb{C}^{\left (M+1 \right ) \times 1}$, where $\mathbb{E}\big \{ \mathbf{s} \mathbf{s}^{H} \big \}= \mathbf{I}$. 
Data streams are then mapped to transmit antennas through a linear precoding matrix $\mathbf{P}=\left [ \mathbf{p}_{c},\mathbf{p}_{1},\cdots \mathbf{p}_{M} \right ]\in \mathbb{C}^{N_{t}\times \left (M+1 \right )}$. This yields a transmit signal $\mathbf{x}\in \mathbb{C}^{N_{t} \times 1}$ given by
\begin{equation}
\mathbf{x} = \mathbf{P}\mathbf{s} = \mathbf{\mathbf{p}}_{c}s_{c} + \sum_{m=1}^{M}\mathbf{\mathbf{p}}_{m}s_{m},
\end{equation}
where $\mathbf{p}_{c} \in \mathbb{C}^{N_{t} \times 1}$ is the common precoder, and $\mathbf{p}_{m}\in \mathbb{C}^{N_{t}\times 1}$ is the $m$-th group's precoder.
Moreover, flexible transmit power constraints are considered in this work, including a total power constraint and per-antenna power constraints. Since the average power of transmit symbols are normalized to be one, the expression of a general transmit power constraint writes as 
\begin{equation}
\mathbf{p}_{c}^{H}\mathbf{D}_{l}\mathbf{p}_{c}+\sum_{m=1}^{M}\mathbf{p}_{m}^{H}\mathbf{D}_{l}\mathbf{p}_{m}\leq P_{l}, \ l \in  \mathcal{L} = \left \{ 1, \cdots,L \right \},
\end{equation}
where  $P_{l}$ is the $l$-th power limit, and $\mathbf{D}_{l}$ is a diagonal shaping matrix changing among different demands. In particular, when the focus is on a total transmit power constraint, let $L =1,\ \mathbf{D}_{l}=\mathbf{I}$ and  $P_{l} = P > 0$, from which $P$ equals to the transmit SNR.
However, in some practical implementations, using individual amplifiers per-antenna causes the lack of flexibility in sharing energy resources.
Such a scenario is typically found in multibeam satellite communications because flexible on-board payloads are costly and complex to implement.
Per-antenna available power constraints are taken into account by setting
$L = N_{t}$, and $P_{l} = P/N_{t}$. The matrix ${D}_{l}$ becomes a zero matrix except its $l$-th diagonal element equaling to 1.

Then, we define $\mu:\mathcal{K}\rightarrow \mathcal{M}$ as mapping a user to its corresponding group. The signal received at user-$k$ can be expanded as
\begin{equation}
y_{k} = \mathbf{h}_{k}^{H} \mathbf{\mathbf{p}}_{c}s_{c} + \mathbf{h}_{k}^{H} \mathbf{\mathbf{p}}_{\mu\left ( k \right ) }s_{\mu\left ( k \right ) } + \mathbf{h}_{k}^{H} \sum^{M}_{ j=1, j\neq \mu\left ( k \right ) }\mathbf{\mathbf{p}}_{j}s_{j}+ n_{k} ,
\end{equation}
where $\mu\left ( k \right )$ is the group index of user-$k$.
Each user firstly decodes the common stream $s_{c}$ and treats $M$ private streams as noise. The SINR of decoding $s_{c}$ at user-$k$ is
\begin{equation}
\gamma_{c,k}=\frac{\left | \mathbf{h}_{k}^{H} \mathbf{p}_{c} \right |^{2}}{ \left |\mathbf{h}_{k}^{H} \mathbf{p}_{\mu\left ( k \right )}   \right |^{2}+ \sum^{M}_{j=1, j\neq \mu\left ( k \right )}\left |\mathbf{h}_{k}^{H} \mathbf{p}_{j}   \right |^{2} +\sigma _{n}^{2}\ }.
\end{equation}
Its corresponding achievable rate writes as $R_{c,k} =\log_{2} \left ( 1+ \gamma_{c,k} \right )$. To guarantee that each user is capable of decoding $s_{c}$, we define a common rate $R_{c}$ at which $s_{c}$ is communicated
\begin{equation}
R_{c}\triangleq \min _{k\in \mathcal{K}} R_{c,k}.
\end{equation}
Note that $s_{c}$ is shared among groups such that $R_{c} \triangleq \sum_{m=1}^{M}C_{m}$, where $C_{m}$ corresponds to group-$m$'s portion of common rate. 
After the common stream $s_{c}$ is decoded and removed through Successive Interference Cancellation (SIC), each user then decodes its desired private stream by treating all the other interference streams as noise.
The SINR of decoding $s_{\mu\left ( k \right )}$ at user-$k$ is given by
\begin{equation}
\gamma_{k}=\frac{\left | \mathbf{h}_{k}^{H} \mathbf{p}_{\mu\left ( k \right )} \right |^{2}}{ \sum^{M}_{j=1, j\neq \mu\left ( k \right )}\left |\mathbf{h}_{k}^{H} \mathbf{p}_{j}   \right |^{2} +\sigma _{n}^{2}\ }.
\end{equation}
Its corresponding achievable rate is $R_{k} =\log_{2} \left ( 1+ \gamma_{k} \right )$. 
In terms of group-$m$, the multicast information $s_{m}$ should be decoded by all users in $\mathcal{G}_{m}$. Thus, the shared information rate $r_{m}$ is determined by the weakest user in $\mathcal{G}_{m}$ and defined as
\begin{equation}
r_{m}\triangleq \min _{i\in \mathcal{G}_{m} }  R_{i}.
\end{equation}
The $m$-th group-rate is composed of $C_{m}$ and $r_{m}$, and writes as
\begin{equation}
r_{g,m}^{RS} = C_{m} + r_{m} = C_{m} + \min _{i\in \mathcal{G}_{m} }  R_{i}.
\end{equation}

In addition, the conventional linear precoding (NoRS) for multigroup multicasting is revisited.
Unlike RS, information intended to each group is encoded directly to a single stream, i.e. $W_{m}\rightarrow s_{m},\ \forall  m \in \left \{ 1\cdots M \right \}$, rather than splitting into a common part and private part. The symbol vector to be transmitted is $\mathbf{s}=\left [ s_{1},\cdots ,s_{M}\right ]^{T}\in \mathbb{C}^{ M \times 1}$, where $\mathbb{E}\left \{ \mathbf{s} \mathbf{s}^{H} = \mathbf{I}\right \}$. 
At user sides, each user decodes its desired stream and treats all the interference streams as noise. 
Following the same multicast logic as (7),
the $m$-th group rate of NoRS writes as
\begin{equation}
r_{g,m}^{NoRS} =  r_{m} \triangleq \min _{i\in \mathcal{G}_{m} }  R_{i}.
\end{equation}
Through the description above, we can observe that RS is a more general scheme\footnote{RS is also a more general framework that encompass NOMA as a special case \cite{joudeh2017rate, clerckx2019rate, mao2018rate, mao2019rate}. Since NOMA leads to a waste of spatial resources and multiplexing gain/ DoF (and therefore rate loss) in multi-antenna settings at the additional expense of large receiver complexity, as demonstrated extensively in \cite{clerckx2019rate, mao2018rate}, we do not compare with NOMA in this work.}
which encompasses NoRS as a special case. RS boils down to NoRS by discarding its common stream and allocating all the transmit power to its private streams.

\textit{Remark 1: The encoding complexity and receiver complexity of RS are slightly higher than NoRS. For the
one-layer RS in a M-group multigoup multicast MISO BC, M + 1 streams need to be encoded in
contrast to M streams for NoRS. One-layer RS requires one SIC at each user sides while NoRS
does not require any SIC.}
\subsection{CSIT Uncertainty and Scaling}
Imperfect CSIT is considered in this work while the channel state information at each receiver (CSIR) is assumed to be perfect.
To model CSIT uncertainty, channel matrix $\mathbf{H}$ is denoted as the sum of a channel estimate
$\widehat{\mathbf{H}} \triangleq \big [\widehat{ \mathbf{h}}_{1}, \cdots,\widehat{\mathbf{h}}_{K} \big ]$
and a CSIT error
$\widetilde{\mathbf{H}} \triangleq \big [\widetilde{ \mathbf{h}}_{1}, \cdots,\widetilde{\mathbf{h}}_{K} \big ]$, i.e.
$\mathbf{H} = \widehat{\mathbf{H}} + \widetilde{\mathbf{H}}$. 
CSIT uncertainty can be characterized by a conditional density
$f_{\mathbf{H}\mid \widehat{\mathbf{H}}}\big ( \mathbf{H} \mid \widehat{\mathbf{H}}  \big ) $ \cite{joudeh2016sum}. 
Taking each channel vector separately, the CSIT error variance $\sigma_{e,k}^{2}\triangleq  \mathbb{E}_{ \widetilde{\mathbf{h}}_{k} }\big \{ \big \| \widetilde{\mathbf{h}}_{k} \big \|^{2} \big \}$ is allowed to decay as $\mathit{O}\left (P ^{-\alpha_{k} } \right )$ \cite{joudeh2016sum,joudeh2016robust,jindal2006mimo,caire2007required},
where $\alpha_{k} \in \left [0,  \infty\right )$ is the scaling factor which quantifies CSIT quality of the $k$-th user. 
Equal scaling factors among users are assumed for simplicity in this model, i.e. $\alpha_{k}=\alpha$.
 For a finite non-zero $\alpha$, CSIT uncertainty decays as $P$ grows, (e.g. by increasing the number of feedback bits).
In extreme cases, $\alpha=0$ corresponds to a non-scaling CSIT, (e.g. with a fixed number of feedback bits). $\alpha \rightarrow \infty $ represents perfect CSIT, (e.g. with infinite number of feedback bits).
The scaling factor is truncated such that 
$\alpha \in\left [ 0,1 \right ]$ in this context since $\alpha=1$ corresponds to perfect CSIT in the DoF sense \cite{joudeh2016sum,joudeh2016robust}.

\section{Max-Min Fair DoF Analysis}
To characterize the performance of RS and NoRS for max-min fair (MMF) multigroup multicasting with imperfect CSIT, MMF-DoF of both schemes are investigated. 
The MMF-DoF, also called MMF multiplexing gain or symmetric multiplexing gain, corresponds to the maximum multiplexing gain that can be simultaneously achieved across multicast groups.
It reflects the pre-log factor of MMF-rate at high SNR.
The larger MMF-DoF is, the faster MMF-rate increases with SNR.
One would therefore like to use communication schemes with the largest possible DoF.
Motivated by mitigating interference at receivers, the beamforming used in this section is sufficient from the DoF perspective
since DoF can be roughly interpreted as the number of interference-free streams simultaneously communicated in a single channel use \cite{joudeh2017rate,joudeh2016sum}.

\subsection{Max-Min Fair DoF of NoRS}
We start from NoRS, and define the $k$-th user-DoF as 
$
{D}_{k} \triangleq \lim_{P\rightarrow \infty}\frac{{R}_{k}\left ( P \right )}{\log_{2}\left ( P \right )}
$.
The $m$-th group-DoF is given by 
$ {d}_{m}^{NoRS} \triangleq 
\lim_{P\rightarrow \infty}\frac{r_{g,m}^{NoRS}\left ( P \right )}{\log_{2}\left ( P \right )}
= \min_{i\in \mathcal{G}_{m}} {D}_{i}$, and
$d^{NoRS} \triangleq \min_{m \in \mathcal{M}} d_{m}^{NoRS}$ is achieved by all groups. 
For a given beamforming $\mathbf{P}=\left [ \mathbf{p}_{1},\cdots \mathbf{p}_{M} \right ]\in \mathbb{C}^{N_{t}\times M}$, $d^{NoRS}$ represents the MMF-DoF.
It interprets the maximum fraction of an interference-free stream that can be simultaneously communicated amongst groups. 
Since each user is equipped with only one antenna, we have
\begin{equation}
d^{NoRS} \leq {d}_{m}^{NoRS} \leq {D}_{i} \leq 1, \quad  \forall \ i\in \mathcal{G}_{m},\ m \in \mathcal{M}.
\end{equation}
\textit{\textbf{Proposition 1.}} The optimum MMF-DoF achieved by NoRS is given by 
\begin{equation}
    d^{*NoRS} =\left\{
\begin{aligned}
 &\alpha,\ N_{t}\geq  K-G_{1}+1
\\
& \frac{\alpha}{2},\ K-G_{M}+1\leq N_{t}< K-G_{1}+1
\\
&0,  \ 1 \leq N_{t}<K-G_{M}+1.
\end{aligned}
\right.
\end{equation}

The achievability of Proposition 1 is discussed as follows by providing at least one feasible beamforming that achieves the DoF in (11).
Next, results in (11) are derived as tight upper-bounds from the converse, which completes the proof of Proposition 1.

\textit{1) Achievability of Proposition 1:}

 To mitigate inter-group interference observed by each user, we aim to design the precoders such that $\widehat{\mathbf{h}}_{k}^{H} \mathbf{p}_{m}= 0$, $\forall \ m \in \mathcal{M},\ k\in \mathcal{K}\setminus\mathcal{G}_{m}$. 
 Define $\widehat{\mathbf{H}}_{m}$ as the composite channel estimate of users in group-$m$, we have $\mathbf{p}_{m} \in \mathrm{null}\big( \widehat{\mathbf{H}}_{\overline{m}}^{H}  \big )$,
where $\widehat{\mathbf{H}}_{\overline{m}} \triangleq \big [\widehat{\mathbf{H}}_{1},\cdots\widehat{\mathbf{H}}_{m-1},\widehat{\mathbf{H}}_{m+1}\cdots ,\widehat{\mathbf{H}}_{M} \big ] \in \mathbb{C}^{N_{t} \times \left (K-G_{m}  \right )}$
 is a channel estimate matrix excluding $\widehat{\mathbf{H}}_{m} $.
 All the channel vectors are assumed to be independent.
 To satisfy $\mathrm{dim}\big ( \mathrm{null}\big ( \widehat{\mathbf{H}}_{\overline{m}}^{H}  \big )  \big )\geq 1$, a minimum number of transmit antennas is required, as follows
 \begin{equation}
N_{t} \geq K-G_{m}+1.
\end{equation}
(12) ensures sufficient $N_{t}$ to place $\mathbf{p}_{m}$ in the null space of its unintended groups.
Primary inter-group interference caused by the $m$-th precoder can be eliminated.
Without loss of generality, group sizes are assumed in an ascending order: $G_{1}\leq G_{2} \leq \cdots\leq G_{M}$. 
In an underloaded scenario, condition (12) has to hold for all $m \in \mathcal{M}$, and we rewrite it as 
 \begin{equation}
N_{t} \geq K-G_{1}+1.
\end{equation}

When (13) is satisfied, the system is underloaded. 
Considering equal power allocation such that $\left \| \mathbf{p}_{1} \right \|^{2} =  \cdots =\left \| \mathbf{p}_{M} \right \|^{2}=\frac{P}{M}$, user-$k$ received signal and the scaling of different received signal components  are expressed by
\begin{equation}
y_{k} = \overbrace{\mathbf{h}_{k}^{H} \mathbf{\mathbf{p}}_{\mu\left ( k \right )}s_{\mu\left ( k \right )}}^{\mathit{O}\left (P \right )} 
+ \overbrace{\widetilde{\mathbf{h}}_{k}^{H} \sum^{M}_{j=1,j\neq \mu\left ( k \right )}\mathbf{\mathbf{p}}_{j}s_{j}}^{\mathit{O}\left (P^{1-\alpha} \right )} 
+   \overbrace{n_{k}}^{\mathit{O}\left (P ^{0}\right )}.
\end{equation}
The second term is named as residual interference
caused by imperfect CSIT.
All the primary inter-group interference $\widehat{\mathbf{h}}_{k}^{H} \sum^{M}_{j=1,j\neq \mu\left ( k \right )}\mathbf{\mathbf{p}}_{j}s_{j}$ 
has been eliminated.
Since the channel state does not depend on $P$, we have 
$\big \| \mathbf{h}_{k} \big \|^{2},\big \| \mathbf{\widehat{h}}_{k} \big \|^{2}=\mathit{O}\left (1 \right )
$.
The residual interference term scales as $\mathit{O}\left (P^{1-\alpha} \right )$, with CSIT error variance decaying as $\mathit{O}\left (P^{-\alpha} \right )$. 
Note that when $\alpha = 1$, the residual interference is reduced to the noise level, and it corresponds to perfect CSIT from the DoF sense.
With $\alpha \in\left [ 0,1 \right ]$,
$\gamma_{k} $ scales as $\mathit{O}\left (P^{\alpha} \right )$, from which
$D_{k}=\alpha $ at each user.
For all $m \in \mathcal{M}$, $d_{m}^{NoRS}=\alpha $,
and hence the MMF-DoF $d^{NoRS} = \alpha$.

When $N_{t} < K-G_{1}+1$, the system becomes overloaded. If reducing the spatial dimensions to $N_{t} < K-G_{M}+1$, 
it is evident that the inter-group interference
caused by each precoder cannot be eliminated. Such scenario is identified as fully-overloaded \cite{joudeh2017rate}, and its MMF-DoF collapses to 0. 
Next, we focus on the partially-overloaded in which $K-G_{M}+1 \leq  N_{t} < K-G_{1}+1$.
We generally assume 
$N_{t} = K-G_{x}+1$, where the group index $x\in \left ( 1,M \right ]$. 
Following the logic of (12), primary inter-group interference caused by the $\left [ x,M \right ]$-th group can be nulled if the precoders are designed such that 
$\mathbf{p}_{m}\in\mathrm{null} \big( \widehat{\mathbf{H}}^{H}_{\overline{m}} \big ),\forall\  m\in  \left [ x,M \right ]$. 
In addition, since $N_{t} = K-G_{x}+1> \left (K-G_{x}  \right )-G_{1}+1$, 
the system excluding group-$x$ can be regarded as underloaded. 
Thus, we design
$\mathbf{p}_{m}\in\mathrm{null} \big ( \widehat{\mathbf{H}}^{H}_{\overline{m},\overline{x}} \big ),\forall\  m\in  \mathcal{M}\setminus x$
to remove inter-group interference among $\mathcal{M}\setminus x$.
The beamforming directions described above can be concluded as

\begin{equation}
\mathbf{p}_{m}\in \left\{
\begin{aligned}
 &\mathrm{null} \big ( \widehat{\mathbf{H}}^{H}_{\overline{m},\overline{x}} \big ),\ \forall\  m\in  \left [ 1,x \right )
\\
& \mathrm{null} \big ( \widehat{\mathbf{H}}^{H}_{\overline{m}} \big ),\ \forall\  m\in  \left [ x,M \right ].
\end{aligned}
\right.
\end{equation}
An example of the power allocation is 
\begin{equation}
\left \| \mathbf{p}_{m} \right \|^{2}= \left\{
\begin{aligned}
 &\frac{P^{\beta}}{M-1},\quad \forall m \in \mathcal{M} \setminus x
\\
& P - P^{\beta}, \quad m \in x.
\end{aligned}
\right.
\end{equation}

where $\beta \in \left [ 0,1 \right ] $ is a power partition factor. 
User-$k$'s received signal is given by 

\begin{equation}
y_{k} =\left\{
\begin{aligned}
 &\overbrace{\mathbf{h}_{k}^{H}\mathbf{p}_{\mu\left ( k \right )}s_{\mu\left ( k \right )} }^{\mathit{O}\left (P ^{\beta}\right )} 
+ \overbrace{\widetilde{\mathbf{h}}_{k}^{H}\sum _{j=1,j\neq\mu\left ( k \right ),j\neq x }^{M}\mathbf{p}_{j}s_{j} }^{\mathit{O}\left (P ^{\beta-\alpha}\right )} \\
& + \overbrace{\widetilde{\mathbf{h}}_{k}^{H} \mathbf{p}_{x}s_{x}}^{\mathit{O}\left (P ^{1-\alpha}\right )} 
 + \overbrace{n_{k}}^{\mathit{O}\left (P ^{0}\right )}
, \  \forall \ k \in \mathcal{K} \setminus \mathcal{G}_{x}.
\\
& \overbrace{\mathbf{h}_{k}^{H}\mathbf{p}_{x}s_{x} }^{\mathit{O}\left (P \right )}
+ \overbrace{\mathbf{h}_{k}^{H}\sum _{j\in \left [ 1,x \right )} \mathbf{p}_{j}s_{j} }^{\mathit{O}\left (P ^{\beta}\right )}
+ \overbrace{\widetilde{\mathbf{h}}_{k}^{H}\sum _{i \in \left ( x,M \right ]}\mathbf{p}_{i}s_{i}}^{\mathit{O}\left (P ^{\beta - \alpha}\right )} \\
& +\overbrace{ n_{k}}^{\mathit{O}\left (P ^{0}\right )}
, \  \forall \ k \in \mathcal{G}_{x}.
\end{aligned}
\right.
\end{equation}

It is observed that $\mathcal{G}_{x}$ bear both residual interference and interference from groups $\left [ 1,x \right )$, while $\mathcal{K} \setminus \mathcal{G}_{x}$ see only residual interference. 
$\gamma_{k}$ at user $k \in \mathcal{K} \setminus \mathcal{G}_{x}$ scales as $\mathit{O}\left (P ^{\beta + \alpha -1}\right )$, and $\gamma_{k}$ at user $\ k \in \mathcal{G}_{x}$ scales as $\mathit{O}\left (P ^{1 - \beta}\right )$. 
Achieving max-min fair DoF requires the same DoF amongst groups.
By setting $\beta = 1- \frac{\alpha }{2}$, all users' SINRs scale as $\mathit{O}\left (P^{\frac{\alpha }{2}} \right )$. It turns out that $d_{m}^{NoRS}=\frac{\alpha }{2}$ for all $m \in \mathcal{M}$, and the MMF-DoF $d^{NoRS} = \frac{\alpha }{2}$ is achieved. Multiplexing gains are partially achieved.
Importantly, such partially-overloaded scenario does not exist when the group sizes are equal.

\textit{2) Converse of Proposition 1:}

Proposition 1 is further shown as a tight upper-bound for any feasible NoRS beamforming. 
Here, we generally assume the power allocation $\left \| \mathbf{p}_{1} \right \|^{2},\cdots , \left \| \mathbf{p}_{M} \right \|^{2}$ scale as $\mathit{O}\left (P ^{a_{1}}\right ),\cdots ,\mathit{O}\left (P ^{a_{M}}\right )$, 
where $a_{1},\cdots ,a_{M} \in \left [ 0,1 \right ] $ are power partition factors.
For each $m \in \mathcal{M}$, $\mathcal{I}_{m}\subset \mathcal{M}$ is defined as a group set with precoding vectors interfering with the $m$-th group, while $\mathcal{R}_{m}\subset \mathcal{M}$ is defined as a group set with precoding vectors that only cause residual interference to the $m$-th group. 
We define $\overline{a}_{m} \triangleq \max_{j \in \mathcal{I}_{m}} a_{j}$, and $\overline{\overline{a}}_{m} \triangleq \max_{j \in \mathcal{R}_{m}} a_{j}$. 
Note that $\overline{a}_{m} = 0$ for $\mathcal{I}_{m}  =  \o $, and $\overline{\overline{a}}_{m} = 0$ for $\mathcal{R}_{m}  =  \o $.
For each $m \in \mathcal{M}$, there exists at least one user $k \in \mathcal{G}_{m}$ with SINR scaling as $\mathit{O}\big (P^{ \min \big \{ \left (a_{m} - \overline{a}_{m}  \right )^{+},\  \left (a_{m} - \overline{\overline{a}}_{m} + \alpha \right )^{+}  \big \}} \big )$, since the received signal can be generally written as
\begin{equation}
y_{k} = \overbrace{\mathbf{h}_{k}^{H} \mathbf{\mathbf{p}}_{\mu\left ( k \right )}s_{\mu\left ( k \right )}}^{\mathit{O}\left (P^{a_{m}} \right )}
+\overbrace{\mathbf{h}_{k}^{H} \sum_{j \in \mathcal{I}_{m}} \mathbf{p}_{j}s_{j}}^{\mathit{O}\left (P ^{\overline{a}_{m}}\right )}
+ \overbrace{\widetilde{\mathbf{h}}_{k}^{H} \sum_{i\in \mathcal{R}_{m}}\mathbf{\mathbf{p}}_{i}s_{i}}^{\mathit{O}\left (P^{\overline{\overline{a}}_{m}} - \alpha\right )} 
+   \overbrace{n_{k}}^{\mathit{O}\left (P ^{0}\right )}.
\end{equation}
According to the definition, we obtain an upper-bound for the achievable group-DoF 
\begin{equation}
{d}_{m}^{NoRS} \leq \min \Big \{ \left (a_{m} - \overline{a}_{m}  \right )^{+},\  \left (a_{m} - \overline{\overline{a}}_{m} + \alpha \right )^{+}  \Big \},
\end{equation}
where $\left ( \cdot  \right )^{+}$ ensures DoF non-negativity. 
The achievable MMF-DoF of NoRS satisfies ${d}^{NoRS} \leq {d}_{m}^{NoRS}$ for all $m \in \mathcal{M}$.
Next, we aim to derive its tight upper-bound ${d}^{*NoRS}$ such that ${d}^{NoRS} \leq {d}^{*NoRS}$ 
for any feasible NoRS beamforming in different network load scenarios.

When the system is underloaded, it is obvious that $\mathcal{I}_{m}  =  \o $ and $\mathcal{R}_{m}  = \mathcal{M} \setminus m $ for all $m \in \mathcal{M}$. 
Accordingly, we have $\overline{a}_{m} = 0$ and $\overline{\overline{a}}_{m} = \max_{j \in \mathcal{M} \setminus m} a_{j}$. (19) can be rewritten as 
\begin{equation}
{d}_{m}^{NoRS} \leq \min \Big \{ a_{m}  ,\  \big (a_{m} - \max_{j \in \mathcal{M} \setminus m} a_{j} + \alpha \big )^{+}  \Big \}.
\end{equation}
From (20), we assume $a_{m} - \max_{j \in \mathcal{M} \setminus m} a_{j} + \alpha  >  0$ because $a_{m} - \max_{j \in \mathcal{M} \setminus m} a_{j} + \alpha \leq 0$ limits ${d}^{*NoRS}$ to 0. 
Then, $\left ( \cdot  \right )^{+}$ can be omitted. 
Since ${d}^{NoRS}$ is upper-bounded by taking the average of any two group-DoFs, we have
\begin{align}
{d}^{NoRS} &\leq \frac{{d}_{1}^{NoRS} + {d}_{2}^{NoRS} }{2} 
\\
& \leq  \frac{1}{2}\min \big \{ a_{1}  ,\ a_{1} - \max_{j \in \mathcal{M} \setminus 1} a_{j} + \alpha   \big \} \notag
\\
 &   + \frac{1}{2}\min \big \{ a_{2}  ,\  a_{2} - \max_{j \in \mathcal{M} \setminus 2} a_{j} + \alpha  \big \} 
 \\
& \leq   \frac{1}{2}\big (\ a_{1} - \max_{j \in \mathcal{M} \setminus 1} a_{j} + \alpha    +   a_{2} - \max_{j \in \mathcal{M} \setminus 2} a_{j} + \alpha     \big ) \\
& \leq \alpha.
\end{align}
(23) follows from the fact that point-wise minimum is upper-bounded by any element in the set. (24) is obtained due to $a_{1} \leq \max_{j \in \mathcal{M} \setminus 2} a_{j} $ and $a_{2} \leq \max_{j \in \mathcal{M} \setminus 1} a_{j}$. 

Next, we focus on the partially-overloaded scenario. It is sufficient to show that ${d}^{NoRS} \leq \frac{\alpha}{2}$ for $N_{t} =  K-G_{1}$, as decreasing the number of antennas does not increase DoF. 
Since $N_{t} < K-G_{1}+1$, $\mathbf{p}_{1}$ leads to interference to at least one group. 
We denote such group index as $m_{1}$. 
In this case, we have 
$\mathcal{I}_{m_{1}}  =  1 $ and $\mathcal{R}_{m_{1}}  = \mathcal{M} \setminus \left \{ 1,m_{1} \right \} $, i.e. $\overline{a}_{m_{1}} = 1$ and $\overline{\overline{a}}_{m_{1}} = \max_{j \in \mathcal{M} \setminus \left \{ 1,m_{1} \right \}} a_{j}$. 
Recalling (19), ${d}_{m_{1}}^{NoRS} $ writes as
\begin{equation}
{d}_{m_1}^{NoRS} \leq \min \Big \{ \big (a_{m_{1}} - a_{1}  \big )^{+},\  \big (a_{m_{1}} - \max_{j \in \mathcal{M} \setminus \left \{ 1,m_{1} \right \}} a_{j} + \alpha \big )^{+}  \Big \}.
\end{equation}
For group-1, it is obvious that $\mathcal{I}_{1}  =  \o $ and $\mathcal{R}_{1}  = \mathcal{M} \setminus 1 $, i.e. 
$\overline{a}_{m_{1}} = 0$ and $\overline{\overline{a}}_{m_{1}} = \max_{j \in \mathcal{M} \setminus 1} a_{j}$. 
Then, we have
\begin{equation}
{d}_{1}^{NoRS} \leq \min \Big \{ a_{1},\  \big (a_{1} - \max_{j \in \mathcal{M} \setminus 1} a_{j} + \alpha \big )^{+}  \Big \}.
\end{equation}
By assuming $a_{m_{1}} - a_{1}>0$ and $a_{1} - \max_{j \in \mathcal{M} \setminus 1} a_{j} + \alpha > 0$, the group-DoF ${d}_{m_{1}}^{*NoRS}$ and ${d}_{1}^{*NoRS}$ are not limited to 0.
$\left ( \cdot  \right )^{+}$ can be omitted in both inequalities. 
Since $a_{1} - \max_{j \in \mathcal{M} \setminus 1} a_{j} + \alpha > 0$ leads to $a_{m_{1}} - a_{1} < a_{m_{1}} - \max_{j \in \mathcal{M} \setminus \left \{ 1,m_{1} \right \}} a_{j} + \alpha$, (25) can be rewritten as ${d}_{m_1}^{NoRS} \leq a_{m_{1}} - a_{1}  $.
Following the same logic as (21), ${d}^{NoRS}$ is upper-bounded by taking the average of ${d}_{1}^{NoRS} $ and ${d}_{m_{1}}^{NoRS} $
\begin{align}
{d}^{NoRS} &\leq \frac{{d}_{1}^{NoRS} + {d}_{m_{1}}^{NoRS} }{2} \\
& \leq  \frac{ \min \big \{ a_{1}  ,\ a_{1} - \max_{j \in \mathcal{M} \setminus 1} a_{j} + \alpha   \big \} + a_{m_{1}} - a_{1}  }{2} \\
& \leq  \frac{  a_{1} - \max_{j \in \mathcal{M} \setminus 1} a_{j} + \alpha    +   a_{m_{1}} - a_{1} }{2} \\
& \leq \frac{\alpha}{2}.
\end{align}
(29) is obtained because point-wise minimum is upper-bounded by any element in the set. (30) is obtained due to $a_{m_{1}} -  \max_{j \in \mathcal{M} \setminus 1} a_{j} \leq 0$.

In a fully-overloaded scenario, it is sufficient to show that ${d}^{NoRS}$ is upper-bounded by 0 for $N_{t} =  K-G_{M}$, as further decreasing $N_{t}$ does not increase DoF.
In this case, we have $N_{t} < K-G_{m}+1$ for all $m \in \mathcal{M}$.
Each $\mathbf{p}_{m}$ causes interference to at least one group. 
Here, we assume $a_{m_{2}} = \max_{m \in \mathcal{M}} a_{m}$. The index of group seeing interference from $\mathbf{p}_{m_{2}}$ is denoted by $m_{3}$.
Thus, $d^{NoRS}$ is upper-bounded by 
\begin{align}
 & d^{NoRS} \leq d_{m_{3}}^{NoRS}  \leq \min \Big \{ \left (a_{m_{3}} - a_{m_{2}}  \right )^{+}, \notag \\
 & \left (a_{m_{3}} - \overline{\overline{a}}_{m_{3}} + \alpha \right )^{+}  \Big \} \leq \left (a_{m_{3}} - a_{m_{2}}  \right )^{+} = 0.
\end{align}
Combining the upper-bounds and achievability derived above, Proposition 1 is proved.
For $\alpha = 1$, such result boils down to the Proposition 1 in \cite{joudeh2017rate} with perfect CSIT. 

\textit{
Remark 2: The basic difference between perfect and imperfect CSIT
scenarios while analysing the DoF of NoRS is the existence of residual interference. 
For example, when we consider perfect CSIT \cite{joudeh2017rate}, $N_{t} \geq K - G_{m} +1$ ensures a sufficient number of transmit antennas to place the $m$-th precoder in the null space of all of its unintended groups. Inter-group interference caused by such precoder can be fully eliminated. 
However, considering imperfect CSIT here, only primary inter-group interference can be eliminated.
At least one form of residual interference still exists. }

\textit{
From the above discussion, when the number of transmit antennas is greater than $K - G_{1} +1$, only residual interference will be seen by each user by controlling the beamforming directions
and power allocation.
Otherwise, the system becomes overloaded.
Through beamforming and power control, the MMF-DoF does not collapse to zero directly as in multi-user
unicast or equal-group multigroup multicast systems.
When $N_{t}$ drops below $K - G_{1} +1$,
$M-1$ groups can be regarded as underloaded, seeing only two forms of residual interference as given in the first equation of (17), while the remaining
one group's received signal subspace is partially sacrificed.
 As a result, a MMF-DoF of $\frac{\alpha }{2}$ is achieved through power control.
When $N_{t}$ drops below $K - G_{M} +1$, each multicast group sees interference from all of its unintended groups. 
The MMF-DoF drops to $0$.
}
\subsection{Max-Min Fair DoF of RS}
In RS scheme, the $m$-th group-DoF writes as
${d}_{m}^{RS} \triangleq \lim_{P\rightarrow \infty}\frac{r_{g,m}^{RS}\left ( P \right )}{\log_{2}\left ( P \right )}
= \min_{i\in \mathcal{G}_{m}} {D}_{i} + d_{c,m}$,
where 
$d_{c,m} \triangleq \lim_{P\rightarrow \infty}\frac{C_{m}\left ( P \right )}{\log_{2}\left ( P \right )}$ is provided by common rate portions. 
$d ^{RS}\triangleq \min_{m \in \mathcal{M}} d_{m}^{RS}$ is the MMF-DoF for a given beamforming $\mathbf{P}=
\left [ \mathbf{p}_{c},\mathbf{p}_{1},\cdots \mathbf{p}_{M} \right ]\in \mathbb{C}^{N_{t}\times \left (M+1 \right )}$.
Obviously, we have
\begin{equation}
d^{RS} \leq {d}_{m}^{RS} \leq {D}_{i} + d_{c,m} \leq 1, \quad  \forall \ i\in \mathcal{G}_{m},\ m \in \mathcal{M}.
\end{equation}
\textit{\textbf{Proposition 2.}} The optimum MMF-DoF achieved by RS is given by
\begin{equation}
d^{*RS} \geq \left\{
\begin{aligned}
 & \frac{1-\alpha }{M} + \alpha,\ N_{t}\geq  K-G_{1}+1.
\\
& \frac{1}{1+M-M_{\mathrm{R}}^{*}},\\
& 1\leq N_{t}< K-G_{1}+1, \ \frac{1}{1+M- M_{\mathrm{R}}^{*}} < \alpha \leq 1.
\\
&\alpha  + \frac{1-\left ( 1+M - M_{\mathrm{R}}^{*} \right )\alpha }{M},\  \\
& 1\leq N_{t}< K-G_{1}+1,\  0\leq \alpha \leq \frac{1}{1+M- M_{\mathrm{R}}^{*}}.
\end{aligned}
\right.
\end{equation}

Note that $M_{\mathrm{R}}^{*} $ is the maximum number of groups which can be regarded as underloaded and served by RS beamforming when the system is overloaded.
The inequality indicates that the results provided here are achievable, yet not necessarily optimum.
The achievability and insight are described as follows.

\textit{1) Achievability of Proposition 2:}

When the system is underloaded, i.e. $N_{t}\geq K-G_{1} +1$, we design $\mathbf{p}_{m} \in \mathrm{null}\big ( \widehat{\mathbf{H}}_{\overline{m}}^{H}  \big )$, which follows the same logic as NoRS. 
The direction of $\mathbf{p}_{c}$ is chosen randomly.
Consider the power allocation such that 
$\left \| \mathbf{p}_{1} \right \|^{2}=\cdots = \left \| \mathbf{p}_{M} \right \|^{2}=\frac{P^{\delta }}{M}$, and
$\left \| \mathbf{p}_{c} \right \|^{2}=P-P^{\delta }$,
where $\delta \in \left [ 0,1 \right ]$ is a power partition factor.
The signal received by user-$k$ writes as
\begin{equation}
y_{k} = \overbrace{\mathbf{h}_{k}^{H} \mathbf{\mathbf{p}}_{c}s_{c} }^{\mathit{O}\left (P \right )}
+ \overbrace{\mathbf{h}_{k}^{H} \mathbf{\mathbf{p}}_{\mu\left ( k \right ) }s_{\mu\left ( k \right ) } }^{\mathit{O}\left (P^{\delta } \right )}
+ \overbrace{\widetilde{\mathbf{h}}_{k}^{H} \sum^{M}_{ j=1, j\neq \mu\left ( k \right ) }\mathbf{\mathbf{p}}_{j}s_{j}}^{\mathit{O}\left (P^{\delta -\alpha } \right )}
+ \overbrace{n_{k} }^{\mathit{O}\left (P^{0 } \right )}.
\end{equation}
It can be observed that $s_{c}$ is firstly decoded at each user with SINR $\gamma_{c,k}$ scaling as $\mathit{O}\left (P^{1-\delta } \right )$. 
The common stream can provide a DoF of $1-\delta$.
Since $R_{c}=\sum_{m=1}^{M}C_{m}$, 
sharing $R_{c}$ equally amongst groups leads to max-min fairness,
and $d_{c,m} = \frac{1-\delta  }{M} $ is achieved by each group.
After removing $s_{c}$, each user then decodes $s_{\mu\left ( k \right ) }$ with $\gamma_{k}$ scaling as $\mathit{O}\left (P^{\min \left \{ \alpha ,\delta  \right \}} \right )$. 
For all $k \in \mathcal{K} $, we have ${D}_{k} = \min \left \{ \alpha ,\delta  \right \}$. Therefore, the MMF-DoF $d^{RS} =\min_{m \in \mathcal{M}} d_{m}^{RS} =  \frac{1-\delta  }{M} + \min \left \{ \alpha ,\delta  \right \}$ can be achieved. By setting $\delta =\alpha $, $d^{RS}$ reaches its maximum value at $\frac{1-\alpha }{M} + \alpha$.

Next, in overloaded scenarios, i.e. $ 1 \leq N_{t} < K- G_{1} +1$,
we consider a special case of RS where groups are divided into two subsets, namely $\mathcal{M}_{\mathrm{R}} \subseteq  \mathcal{M}$ and $\mathcal{M}_{\mathrm{C}}=\mathcal{M} \setminus \mathcal{M}_{\mathrm{R}}$. 
Specifically, $\mathcal{M}_{\mathrm{R}}$ is a subset which can be treated as underloaded and served by RS beamforming, while $\mathcal{M}_{\mathrm{C}}$ are the remaining groups and served by degraded beamforming. 
Based on this mixed scheme, messages are split such that 
$W_{m}\rightarrow \left \{ W_{m,c} , W_{m,p} \right \}$ for all $m \in \mathcal{M}_{\mathrm{R}}$, and 
$W_{m}\rightarrow \left \{ W_{m,c} \right \}$ for all $m \in \mathcal{M}_{\mathrm{C}}$. 
Such scheme leads to $\left \| \mathbf{p}_m \right \|^{2}=0$ for all $m \in \mathcal{M}_{\mathrm{C}}$. 
The size of $\mathcal{M}_{\mathrm{R}}$ and $\mathcal{M}_{\mathrm{C}}$ are denoted by
$M_{\mathrm{R}} = \left |\mathcal{M}_{\mathrm{R}}  \right |$
and $M_{\mathrm{C}} = \left |\mathcal{M}_{\mathrm{C}}  \right | = M- M_{\mathrm{R}}$ respectively.
To gain insight into the subset partition, we define 
\begin{equation}
N_{L} =\left\{
\begin{aligned}
 &K -G_{1} - \sum_{j = L+1}^{M}G_{j} +1,\ L \in \left \{ 1,\cdots ,M-1 \right \}
\\
& K-G_{1} +1 ,\ L=M.
\end{aligned}
\right.
\end{equation}
According to (13), $N_{L}$ is the minimum number of transmitting antennas required to regard groups $\left \{ 1,\cdots ,L \right \}$ as underloaded while disregarding all the remaining groups. 
Conversely, if $N_{t}$ satisfies $N_{L}\leq N_{t}< N_{L+1}$, $L$ is interpreted as the maximum number of $M_{\mathrm{R}}$. We can define it as
\begin{equation}
M_{\mathrm{R}}^{*}=
\left\{
\begin{aligned}
 &M , \ N_{t}\geq  N_{M}
\\
& L, \  N_{L}\leq  N_{t}<  N_{L+1},\  \forall L \in \left \{ 1, \cdots, M-1 \right \}.
\end{aligned}
\right.
\end{equation}
For all $m \in \mathcal{M}_{\mathrm{R}}$, beamforming directions are designed as $\mathbf{p}_{m} \in \mathrm{null}\big ( \widehat{\mathbf{H}}_{ \left \{ \overline{m},\overline{\mathcal{M}_{\mathrm{C}}} \right \}
}^{H}  \big )$. $\mathbf{p}_{c}$'s direction is set randomly. Consider the power allocation $\left \| \mathbf{p}_{m} \right \|^{2}=\frac{P^{\delta }}{M_{\mathrm{R}}}$ for all $m \in \mathcal{M}_{\mathrm{R}}$,
and $\left \| \mathbf{p}_{c} \right \|^{2}=P-P^{\delta }$,
where $\delta \in \left [ 0,1 \right ]$. User-k's received signal writes as
\begin{equation}
y_{k} =\left\{
\begin{aligned}
 &\overbrace{\mathbf{h}_{k}^{H}\mathbf{p}_{c}s_{c}}^{\mathit{O}\left (P \right )} + 
\overbrace{\mathbf{h}_{k}^{H}\mathbf{p}_{\mu\left ( k \right )}s_{\mu\left ( k \right )}}^{\mathit{O}\left (P^{\delta } \right )} 
+ \overbrace{\widetilde{\mathbf{h}}_{k}^{H} \sum_{j \in \mathcal{M}_{\mathrm{R}} \setminus \mu\left ( k \right )}\mathbf{p}_{j}s_{j}} ^{\mathit{O}\left (P^{\delta -\alpha } \right )} \\
& +
\overbrace{\sigma _{n}^{2} } ^{\mathit{O}\left (P^{0 } \right )}
, \  \forall \ k \in \left \{ \mathcal{G}_{m} \mid m \in \mathcal{M}_{\mathrm{R}}\right \}.
\\
& \overbrace{\mathbf{h}_{k}^{H}\mathbf{p}_{c}s_{c}}^{\mathit{O}\left (P \right )} + \overbrace{\mathbf{h}_{k}^{H}\sum_{j \in \mathcal{M}_{\mathrm{R}} }\mathbf{p}_{j}s_{j}}^{\mathit{O}\left (P^{\delta } \right )}+
\overbrace{\sigma _{n}^{2}}^{\mathit{O}\left (P^{0 } \right )}
, \\
& \forall \ k \in \left \{ \mathcal{G}_{m} \mid m \in \mathcal{M}_{\mathrm{C}}\right \}.
\end{aligned}
\right.
\end{equation}
Firstly, $s_{c}$ is decoded at each user by treating all the other streams as noise. $\gamma _{c,k}$ is observed to scale as $\mathit{O}\left (P^{1-\delta } \right )$ for $k \in \mathcal{K}$.
Hence, the common stream achieves a DoF of $1-\delta$. 
Since the common rate $R_{c}=\sum_{m=1}^{M}C_{m}$ is divided amongst $\mathcal{M}_{\mathrm{R}}$ and $\mathcal{M}_{\mathrm{C}}$, 
we introduce a fraction $z \in \left [ 0,1 \right ]$ of the common rate such that 
$\sum _{m\in \mathcal{M}_{\mathrm{R}}}C_{m} = z R_{c}$, and $\sum _{m\in \mathcal{M}_{\mathrm{C}}}C_{m} = \left ( 1-z \right )R_{c}$ .
This leads to $d_{c,m} = \frac{z\left ( 1-\delta   \right )}{M_{\mathrm{R}}}$ for $m \in \mathcal{M}_{\mathrm{R}}$ 
and $d_{c,m} = \frac{\left (1-z  \right )\left ( 1-\delta   \right )}{M-M_{\mathrm{\mathrm{R}}}}$ for $m \in \mathcal{M}_{\mathrm{C}}$.
After removing $s_{c}$ through SIC, it can be seen that $\gamma _{k}$ scales as $\mathit{O}\left (P^{\min\left \{ \alpha ,\delta  \right \}} \right )$ in the first subset $\mathcal{M}_{\mathrm{R}}$.
Hence, we have ${D}_{k} = \min \left \{ \alpha ,\delta  \right \}$ for all $k \in \left \{ \mathcal{G}_{m} \mid m \in \mathcal{M}_{\mathrm{R}}\right \}$.
The group-DoF $d^{RS}_{m}$ is given by
\begin{equation}
d^{RS}_{m} =\left\{
\begin{aligned}
 &\frac{z\left ( 1-\delta   \right )}{M_{\mathrm{R}}} + \min\left \{ \alpha ,\delta  \right \} , \quad \forall \ m\in \mathcal{M}_{\mathrm{R}}
\\
& \frac{\left (1-z  \right )\left ( 1-\delta   \right )}{M- M_{\mathrm{\mathrm{R}}}} , \quad \forall \ m\in \mathcal{M}_{\mathrm{C}}.
\\
\end{aligned}
\right.
\end{equation}
To achieve max-min fairness, equal group-DoFs between $\mathcal{M}_{\mathrm{R}}$ and $\mathcal{M}_{\mathrm{C}}$ are required.
On one hand, we assume $\delta \geq \alpha$, and the equation can be written as 
\begin{equation}
\frac{z\left ( 1-\delta   \right )}{M_{\mathrm{R}}} +  \alpha = \frac{\left (1-z  \right )\left ( 1-\delta   \right )}{M- M_{\mathrm{\mathrm{R}}}} .
\end{equation}
Note that there are two variables $\delta$ and $z$ on both sides of (39). Since the two variables cannot be solved simultaneously, we fix one variable to maximize at least one side of (39) while reserving the other variable on both sides. For example, let $\delta = \alpha$ in this case, and then calculate the remaining variable $z$ according to 
\begin{equation}
\frac{z\left ( 1-\alpha   \right )}{M_{\mathrm{R}}} +  \alpha = \frac{\left (1-z  \right )\left ( 1-\alpha   \right )}{M- M_{\mathrm{\mathrm{R}}}} .
\end{equation}
$z = \frac{\left [1-\left ( 1+ M-M_{\mathrm{R}} \right )\alpha  \right ] M_{\mathrm{R}}}{\left (1-\alpha  \right )M }$ is obtained. 
Substitute it into arbitrary side of (40), and the group-DoF $d^{RS}_{m} = \alpha  + \frac{1-\left ( 1+M - M_{\mathrm{R}} \right )\alpha }{M}$ for all $m \in \mathcal{M}$ is derived. 
Moreover, a corresponding condition $0\leq \alpha \leq \frac{1}{1+M- M_{\mathrm{R}}}$ is obtained by considering $0\leq z = \frac{\left [1-\left ( 1+ M-M_{\mathrm{R}} \right )\alpha  \right ] M_{\mathrm{R}}}{\left (1-\alpha  \right )M } \leq 1$.
The MMF-DoF is achieved as $d^{RS} =\min_{m \in \mathcal{M}} d_{m}^{RS} = \alpha  + \frac{1-\left ( 1+M - M_{\mathrm{R}}^{*} \right )\alpha }{M}$, when $0\leq \alpha \leq \frac{1}{1+M- M_{\mathrm{R}}^{*}}$ .

On the other hand, we assume $\delta < \alpha$. The equation in (39) is rewritten as
\begin{equation}
\frac{z\left ( 1-\delta   \right )}{M_{\mathrm{R}}} +  \delta = \frac{\left (1-z  \right )\left ( 1-\delta   \right )}{M- M_{\mathrm{\mathrm{R}}}} .
\end{equation}
There are still two variables $\delta$ and $z$ in (41). In this case, we can set $z=0$ to maximize the right side of (41) and calculate $\delta$ according to 
\begin{equation}
  \delta = \frac{1-\delta}{M- M_{\mathrm{\mathrm{R}}}} .
\end{equation}
By substituting the solution $\delta =\frac{1}{1+M-M_{\mathrm{R}}}$ into arbitrary side of (42), the group-DoF $d^{RS}_{m} = \frac{1}{1+M-M_{\mathrm{R}}}$ for all $m \in \mathcal{M}$ is derived.
Since $\delta =\frac{1}{1+M-M_{\mathrm{R}}} < \alpha$, we obtain the corresponding condition $ \frac{1}{1+M- M_{\mathrm{R}}} < \alpha \leq 1$ for this case. 
Above all, the achievable MMF-DoF of RS is summarized on the right side of Proposition 2.
When $\alpha = 1$, such result boils down to the achievability of Proposition 3 in \cite{joudeh2017rate} with perfect CSIT. In overloaded scenarios, it is noteworthy that the $d^{RS}$ with $\frac{1}{1+M- M_{\mathrm{R}}^{*}} < \alpha \leq 1$ is not a function of $\alpha$ and is the same as that achieved with perfect CSIT. Thus, one can relax the CSIT quality up to $\frac{1}{1+M- M_{\mathrm{R}}^{*}}$ without affecting the MMF-DoF.
However, in the other case when $0\leq \alpha \leq \frac{1}{1+M- M_{\mathrm{R}}^{*}}$, $d^{RS}$ diminishes as the CSIT quality reduces. 

\textit{2) Insight:}

From (37), the interference seen by each user $\ k \in \left \{ \mathcal{G}_{m} \mid m \in \mathcal{M}_{\mathrm{R}}\right \}$ after SIC 
scales as $\mathit{O}\left (P^{\delta -\alpha } \right )$.
As discussed above, we have two
assumptions, namely $\delta \geq \alpha$ and $\delta < \alpha$.
When $\delta \geq \alpha$, this residual interference cannot be ignored.
By setting the power partition factor $\delta\rightarrow \alpha $, we can reduce it to the noise level
and at the same time increase $\gamma _{c,k}$ which scales as $\mathit{O}\left (P^{1-\delta } \right )$ for all $k \in \mathcal{K}$.
To achieve max-min fairness, the common rate factor $z$ is then managed to obtain equal group-DoFs among groups in $\mathcal{M}_{R}$ and $\mathcal{M}_{C}$.
$0\leq \alpha \leq \frac{1}{1+M- M_{\mathrm{R}}^{*}}$ is derived as a corresponding range of this case.
 The MMF-DoF reduces as $\alpha$ goes down.
 Otherwise, when $\delta < \alpha$, such interference is always at the noise level. 
By setting $z\rightarrow 0$, all the common rate $R_{c}$ contributes to $C_{m}$, for all $m \in \mathcal{M}_{C}$. 
The RS scheme used by $\mathcal{M}_{R}$ boils down to NoRS. 
Meanwhile, the group-DoFs of all $  m \in \mathcal{M}_{\mathrm{C}}$ are maximized.
Then, we further manage the power partition factor $\delta$ to achieve max-min fairness amongst all groups. 
$\frac{1}{1+M- M_{\mathrm{R}}^{*}} < \alpha \leq 1$ is derived as the corresponding range. 
In this case, changing $\alpha$ will no longer affect MMF-DoF because the interference
seen by each user $\ k \in \left \{ \mathcal{G}_{m} \mid m \in \mathcal{M}_{\mathrm{R}}\right \}$ after SIC is always at the noise level. 
The MMF-DoF performance remains the same as that achieved with perfect CSIT. 
Such behavior is not observed in partially-overloaded NoRS. 
It can be observed in (17) that the power of interference seen by each user $k \in \mathcal{K} \setminus \mathcal{G}_{x}$ and $k \in \mathcal{G}_{x}$ scales as $\mathit{O}\left (P ^{1 - \alpha }\right )$ and $\mathit{O}\left (P ^{\beta }\right )$ respectively. $\alpha$ will always affect MMF-DoF as $\mathit{O}\left (P ^{1 - \alpha }\right )$ cannot be ignored unless considering perfect CSIT.
To get more insight into the gains provided by RS over NoRS, we substitute (36) into (33) and yield (43).
\begin{equation}
 d^{*RS} \geq \left\{
\begin{aligned}
 & \frac{1-\alpha }{M} + \alpha,\ N_{t}\geq  N_{M}
\\
& \frac{1}{2},\ N_{M-1}\leq N_{t}< N_{M},\ \frac{1}{2} < \alpha \leq 1
\\
&\alpha  + \frac{1-2 \alpha }{M},\  N_{M-1}\leq N_{t}< N_{M},\  0\leq \alpha \leq \frac{1}{2}
\\
&\vdots 
\\
& \frac{1}{M-1},\  N_{2}\leq N_{t}< N_{3},\ \frac{1}{M-1} < \alpha \leq 1
\\
&\alpha  + \frac{1-\left (M-1  \right ) \alpha }{M},\\
& N_{2}\leq N_{t}< N_{3},\  0\leq \alpha \leq \frac{1}{M-1}
\\
& \frac{1}{M},\  1\leq N_{t}< N_{2},\ \frac{1}{M} < \alpha \leq 1
\\
& \frac{1}{M},\  1\leq N_{t}< N_{2},\ 0\leq \alpha \leq \frac{1}{M}
\end{aligned}
\right.                              
\end{equation}            

\begin{table*}
\caption{Achievable MMF-DoF of different strategies}
\label{table_example}
\centering
\begin{threeparttable}[b]
\begin{tabular}{|c|c | c|c  |c|}
\hline

& \multicolumn{2}{c|}{\textbf{Perfect CSIT [16]}} & \multicolumn{2}{c|}{\textbf{Imperfect CSIT [this paper]}} \\ \hline

Strategy & NoRS & RS& NoRS & RS
 \\\hline
$N_{t} \geq N_{M}$ & $1$ & $1$ & $\alpha$ & $\frac{1-\alpha }{M} + \alpha$
\\\hline

$N_{M-1}+G_{1}\leq N_{t}< N_{M}$\tnote{1}
& $\frac{1}{2}$&$\frac{1}{2}$ &$\frac{\alpha }{2}$&

$
\left\{
\begin{aligned}
& \frac{1}{2},\ \frac{1}{2} < \alpha \leq 1
\\

&\alpha  + \frac{1-2 \alpha }{M},\  0\leq \alpha \leq \frac{1}{2}
\\
\end{aligned}
\right.  
$

\\\hline

 $N_{M-1}\leq N_{t}< N_{M-1}+G_{1}$ &0&
 $\frac{1}{2}$ &0&
 $
\left\{
\begin{aligned}
& \frac{1}{2},\ \frac{1}{2} < \alpha \leq 1
\\
&\alpha  + \frac{1-2 \alpha }{M},\  0\leq \alpha \leq \frac{1}{2}
\\
\end{aligned}
\right.  
$
 \\\hline

$N_{M-2}\leq N_{t}< N_{M-1}$ & 0 & $\frac{1}{3}$ &0&
 $
\left\{
\begin{aligned}
& \frac{1}{3},\ \frac{1}{3} < \alpha \leq 1
\\
&\alpha  + \frac{1-3 \alpha }{M},\  0\leq \alpha \leq \frac{1}{3}
\\
\end{aligned}
\right.  
$
 \\\hline
 \vdots & & & &\vdots 
 \\
 
$1\leq N_{t}< N_{2}$ & 0& $\frac{1}{M}$ & 0 & $\frac{1}{M}$\\
\hline

\end{tabular}
\begin{tablenotes}
\item[1] The second line of this table (partially-overloaded scenario) does not exist when the group sizes are equal.
When $N_{t}$ drops below $K-G+1$, $d^{*NoRS}$ decreases to $0$ directly.
\end{tablenotes}

\end{threeparttable}
\end{table*}

By comparing (43) with (11), we can see that the achievable MMF-DoF of RS is always superior than $d^{*NoRS}$, and hence $d^{*RS} \geq d^{*NoRS}$ is guaranteed. 
The gain of RS over NoRS is $\frac{1-\alpha }{M}$ when the system is underloaded.
Once $N_{t}\geq  N_{M}$ is violated, the range of partially-overloaded NoRS $K-G_{M}+1\leq N_{t}< K-G_{1}+1$, (i.e. $N_{M-1} + G_{1}\leq N_{t}< N_{M}$) locates within the range $N_{M-1} \leq N_{t}< N_{M}$. 
For any $0\leq \alpha \leq 1$, the achievable MMF-DoF of RS is still greater than NoRS.
Once $N_{t}$ drops below $N_{M-1} + G_{1}$, by taking $N_{t} = N_{M-1} $ as an example,
the number of antennas is not sufficient to eliminate any inter-group interference through NoRS beamforming. $d^{*NoRS}$ collapses to 0. 
For RS, $d^{*RS}$ is kept by exploiting all the $M_{\mathrm{R}}^{*}$ streams and transmitting the remaining stream through degraded beamforming. This is carried on until RS reducing to a single-stream degraded beamforming. A single DoF is split amongst all the groups. Therefore, $d^{*RS} \geq \frac{1}{M}> 0$ is guaranteed.
For the particular case where all the group size are equal, (i.e. $G_{m}=G,\ \forall m \in \mathcal{M}$), there is not partially-overloaded scenario in (11). When $N_{t}$ drops below $K-G+1$, $d^{*NoRS}$ decreases from $\alpha$ to $0$ directly. However, the expression of $d^{*RS}$ remains the same as (43), which is always greater than $\frac{1}{M}$.

\textit{
Remark 3: The obtained MMF-DoFs of different strategies are listed in Table I, where the first row represents
underloaded and the others are results of overloaded systems. 
}

\textit{
From the above discussion, the MMF-DoF analysis in the underloaded regime is similar for RS and NoRS. Each user sees only residual interference by managing the beamforming directions and power allocation. 
A gain of $\frac{1-\alpha }{M}$ is obtained. 
Thus, we can conclude that in the presence of imperfect CSIT, there is a MMF-DoF gain of RS over NoRS when the system is underloaded. 
This contrasts with perfect CSIT scenarios where both underloaded NoRS and RS can achieve full MMF-DoF of $1$.
Overloaded RS is more challenging since both residual interference and group partitioning method should be considered.
\cite{joudeh2017rate} considers a special case where the groups are partitioned into two subsets, namely $\mathcal{M}_{D}\subseteq \mathcal{M}$ which are served using NoRS, and $\mathcal{M}_{C}\subseteq \mathcal{M} \setminus \mathcal{M}_{D}$ served by degraded beamforming.
The number of groups in $\mathcal{M}_{D}$ is set as the maximum number of groups that can be served by interference-free NoRS (i.e. achieving a group-DoF of $1$ each). 
However, in this work considering imperfect CSIT, NoRS can no longer reach a MMF-DoF of $1$. As shown in Table 1, the maximum achievable MMF-DoF is $\alpha$ when the system is underloaded, while RS outperforms NoRS slightly.
Thus, we consider a different subset partitioning in this work where the groups are divided into $\mathcal{M}_{R}\subseteq \mathcal{M}$ and $\mathcal{M}_{C}\subseteq \mathcal{M} \setminus \mathcal{M}_{R}$.
The number of groups in $\mathcal{M}_{R}$ is chosen as the maximum number of groups which can be served by RS and achieve a MMF-DoF of $\frac{1-\alpha }{M} + \alpha$.
$\mathcal{M}_{C}$ is still served by degraded beamforming.
Accordingly, from the results summarised in Table I, RS is shown to provide MMF-DoF gains and outperform NoRS in overloaded systems.
}


All the discussions above motivate the use of RS over NoRS from a DoF perspective. However, DoF is an asymptotically high SNR metric.  It remains to be seen whether the DoF gain translate into rate gains. To that end, the design of RS for rate maximization at finite SNR needs to be investigated. 
Beamforming schemes that achieve Proposition 1 and Proposition 2 are not necessarily optimum from a MMF-rate sense. 
Therefore, the beamforming directions, power allocation and rate partition can be elaborated by formulating MMF-rate optimization problems as we see in the next section.
Importantly, the DoF analysis provides fundamental grounds, helps drawing insights into the performance limits of various strategies and guide the design of efficient strategy (rate-splitting in this case).

\section{Optimization and Solution}
In this section, an optimization problem is formulated to achieve MMF among multiple co-channel multicast groups subject to a flexible power constraint with imperfect CSIT. 
MMF Ergodic Rate is the metric for both RS and NoRS. 
It reflects long-term performance over varying channel states. 
Given a long sequence of channel estimates, the MMF Ergodic Rate (ER) can be measured by updating precoders based on each short-term MMF Average Rate (AR). 
To begin with, the relationship between AR and ER is introduced.
ARs of user-$k$ are short-term measures defined as
\begin{subequations}
\begin{align}
\overline{R}_{c,k}\big ( \widehat{\mathbf{H}} \big ) &\triangleq \mathbb{E}_{\mathbf{H} \mid \widehat{\mathbf{H}}}\Big \{ R_{c,k} \big ( \mathbf{H}, \widehat{\mathbf{H}} \big )  \mid  \widehat{\mathbf{H}}\Big \} \tag{44a}
\\
\overline{R}_{k}\big ( \widehat{\mathbf{H}} \big ) &\triangleq \mathbb{E}_{\mathbf{H} \mid \widehat{\mathbf{H}}}\Big \{ R_{k} \big ( \mathbf{H}, \widehat{\mathbf{H}} \big )  \mid  \widehat{\mathbf{H}}\Big \}.  \tag{44b}
\end{align}
\end{subequations}
Here, ARs have captured the expected performance over CSIT error distribution for a given channel state estimation, where $R_{c,k}\big ( \mathbf{H},\widehat{\mathbf{H}} \big )$ and $R_{k}\big ( \mathbf{H},\widehat{\mathbf{H}} \big )$ are determined by $\big \{ \mathbf{H},\widehat{\mathbf{H}} \big \}$.
According to the law of total expectation, the ERs of user-$k$ can be expressed by
\begin{subequations}
\begin{align}
\mathbb{E}_{\left \{ \mathbf{H},\widehat{\mathbf{H}} \right \}}\Big \{ R_{c,k}\big ( \mathbf{H},\widehat{\mathbf{H}}  \big ) \Big \}
&= \mathbb{E}_{\widehat{\mathbf{H}}}\left \{ \mathbb{E}_{\left \{ \mathbf{H\mid \widehat{\mathbf{H}}} \right \} } 
 \left \{  R_{c,k}\big ( \mathbf{H},\widehat{\mathbf{H}}  \big ) \mid \widehat{\mathbf{H}}\right \}\right \} \notag \\ 
 &\triangleq \mathbb{E}_{\widehat{\mathbf{H}}}\left \{ \bar{R} _{c,k} \big ( \widehat{\mathbf{H} }\big )\right \} \tag{45a}
\\
\mathbb{E}_{\left \{ \mathbf{H},\widehat{\mathbf{H}} \right \}}\left \{ R_{k}\big ( \mathbf{H},\widehat{\mathbf{H}} \big ) \right \}
& = \mathbb{E}_{\widehat{\mathbf{H}}}\left \{ \mathbb{E}_{\left \{ \mathbf{H\mid \widehat{\mathbf{H}}} \right \} } 
 \left \{  R_{k}\big ( \mathbf{H},\widehat{\mathbf{H}}  \big ) \mid \widehat{\mathbf{H}}\right \}\right \}  \notag \\ 
 &\triangleq \mathbb{E}_{\widehat{\mathbf{H}}}\left \{ \bar{R} _{k} \big ( \widehat{\mathbf{H} }\big )\right \}. \tag{45b}
\end{align}
\end{subequations}
It turns out that measuring ERs is transformed into measuring ARs over the variation of $\widehat{\mathbf{H}} $.
Therefore, the MMF ER of RS can be characterized by $\mathbb{E}_{\widehat{\mathbf{H}}}\left \{ \mathcal{F}_{RS}\left ( P \right ) \right \}$, where $\mathcal{F}_{RS}\left ( P \right )$ is
a stochastic problem of achieving MMF AR for a given channel estimate $\widehat{\mathbf{H}} $.
\begin{align}
\mathcal{F}_{RS}\left ( P \right ): \quad
&
\max _{\mathbf{\overline{c}},\mathbf{P}} \min_{m\in \mathcal{M}} \big ( \overline{C}_{m} + \min _{i\in \mathcal{G}_{m} }  \overline{R}_{i}\big )
\\
s.t. \quad
&\overline{R}_{c,k}\geq \sum_{m=1}^{M} \overline{C}_{m},\  \forall k \in \mathcal{K}
\\
&\overline{C}_{m} \geq 0, \  \forall m \in \mathcal{M}
\\
&\mathbf{p}_{c}^{H}\mathbf{D}_{l}\mathbf{p}_{c}+\sum_{m=1}^{M}\mathbf{p}_{m}^{H}\mathbf{D}_{l}\mathbf{p}_{m}\leq P_{l},\ l \in \mathcal{L}
\end{align}
By solving $\mathcal{F}_{RS}\left ( P \right )$, 
$\mathbf{\overline{c}} \triangleq \left [ \overline{C}_{1},\cdots ,\overline{C}_{M} \right ]$ 
and $\mathbf{P} = \left [ \mathbf{p}_{c},\mathbf{p}_{1},\cdots \mathbf{p}_{M} \right ]$ are jointly optimized.
Since the average common rate is given by $\overline{R}_{c} = \sum_{m=1}^{M}\overline{C}_{m}=\min _{k \in \mathcal{K}} \overline{R}_{c,k}$, we use constraint (47) to ensure $s_{c}$ is decoded by each user. 
Constraint (48) implies that each portion of $\overline{R}_{c}$ is non-negative and (49) is the transmit power constraint described in Section II. 

 Similarly, the corresponding stochastic problem in NoRS version is formulated as
 \begin{align}
\mathcal{F}_{NoRS}\left ( P \right ): \quad
&
\max _{\mathbf{P}} \min_{m\in \mathcal{M}} \big (  \min _{i\in \mathcal{G}_{m} }  \overline{R}_{i}\big)
\\
s.t. \quad
&\sum_{m=1}^{M}\mathbf{p}_{m}^{H}\mathbf{D}_{l}\mathbf{p}_{m}\leq P_{l}, \ l \in \mathcal{L} 
\end{align}
where $\mathbf{P}=\left [\mathbf{p}_{1},\cdots, \mathbf{p}_{M} \right ]$ is optimized to solve $\mathcal{F}_{NoRS}\left ( P \right )$. (51) is the flexible transmit power constraint. 
NoRS is a sub-scheme of RS by discarding the common stream. 
Solving $\mathcal{F}_{NoRS}\left ( P \right )$ is a special case of $\mathcal{F}_{RS}\left ( P \right )$ by fixing $\mathbf{\overline{c}}=0$ and $\left \| \mathbf{p}_{c} \right \|^{2} = 0$. 
We will focus on solving the RS-based problem in the following discussion.

Sample Average Approximation (SAA) is then adopted to convert $\mathcal{F}_{RS}\left ( P \right )$ into a deterministic problem denoted by $\mathcal{F}_{RS}^{\left ( S \right )}\left ( P \right )$. 
 For a given  $\widehat{\mathbf{H}}$ and sample index set $\mathfrak{S} \triangleq \left \{ 1,\cdots,S \right \}$, we construct a realization sample $\mathbb{H}^{\left ( S \right )} \triangleq \big \{ \mathbf{H}^{\left (s  \right )}= \widehat{\mathbf{H}}+ \widetilde{\mathbf{H}}^{\left (s  \right )}\mid \widehat{\mathbf{H}} , s \in \mathfrak{S}\big \} $ containing $S$ i.i.d realizations drawn from a conditional distribution with density $f_{\mathbf{H}\mid \widehat{\mathbf{H}}}\big ( \mathbf{H} \mid \widehat{\mathbf{H}}  \big ) $.
These realizations are available at the transmitter and can be used to approximate the ARs experienced by each user through Sample Average Functions (SAFs). When $S\rightarrow \infty$, according to the strong law of large numbers, we have
 \begin{align}
\overline{R}_{c,k} &= \lim_{S\rightarrow \infty} \overline{R}_{c,k}^{\left (S  \right )} =\lim_{S\rightarrow \infty} \frac{1}{S}\sum_{s=1}^{S} R_{c,k}\big ( \mathbf{H}^{\left (s  \right )} \big )
\\
 \overline{R}_{k} &= \lim_{S\rightarrow \infty}\overline{R}_{k}^{\left (S  \right )} =\lim_{S\rightarrow \infty} \frac{1}{S}\sum_{s=1}^{S} R_{k}\big ( \mathbf{H}^{\left (s  \right )} \big )
 \end{align}
 where $R_{c,k}\left ( \mathbf{H}^{\left (s  \right )} \right )$ and $R_{k}\left ( \mathbf{H}^{\left (s  \right )} \right ), \ s \in \mathfrak{S}$ are the rates based on the realization sample $\mathbf{H}^{\left (s  \right )}$.
Accordingly, the deterministic problem can be written as
\begin{align}
\mathcal{F}_{RS}^{\left ( S \right )}\left ( P \right ): \quad
&
\max _{\mathbf{\overline{c}},\mathbf{P}} \min_{m\in \mathcal{M}} \big ( \overline{C}_{m} + \min _{i\in \mathcal{G}_{m} }  \overline{R}_{i}^{\left ( S \right )}\big )
\\
s.t. \quad
&\overline{R}_{c,k}^{\left ( S \right )}\geq \sum_{m=1}^{M} \overline{C}_{m},\  \forall k \in \mathcal{K}
\\
&\overline{C}_{m} \geq 0, \  \forall m \in \mathcal{M}
\\
&\mathbf{p}_{c}^{H}\mathbf{D}_{l}\mathbf{p}_{c}+\sum_{m=1}^{M}\mathbf{p}_{m}^{H}\mathbf{D}_{l}\mathbf{p}_{m}\leq P_{l}, \ l \in \mathcal{L}
\end{align}

 Note that $\mathcal{F}_{RS}^{\left ( S \right )}\left ( P \right )$ is a non-convex optimization problem which is very challenging to solve. 
 
 The WMMSE approach, initially proposed in \cite{christensen2008weighted}, is effective in solving problems containing non-convex superimposed rate expressions, i.e. RS problems.
 In this work, we use AO (alternating optimization) and a modified WMMSE approach to solve the above problem.
 A deterministic SAF version of the Rate-WMMSE relationship relationship is constructed in \cite{9145200} such that
 \begin{subequations}
 \begin{align}
\overline{\xi}_{c,k}^{MMSE\left (S \right )} &= \min_{\mathbf{g}_{c,k},\mathbf{u}_{c,k}}\overline{\xi} _{c,k}^{\left ( S \right )}= 1- \overline{R}_{c,k}^{\left ( S \right )} \tag{58a}
\\
\overline{\xi}_{k}^{MMSE\left (S  \right )} &= \min_{\mathbf{g}_{k},\mathbf{u}_{k}}\overline{\xi} _{k}^{\left ( S \right )}= 1- \overline{R}_{k}^{\left ( S \right )}.\tag{58b}
\end{align}
\end{subequations}
This relationship holds for the whole set of stationary points \cite{joudeh2016sum}.
For a given channel estimate, $\overline{\xi}_{c,k}^{MMSE\left (S  \right )}$ and $\overline{\xi}_{k}^{MMSE\left (S  \right )}$ represent average WMMSEs approximated by the SAFs. We have $\overline{\xi}_{c,k}^{MMSE\left (S \right )}=\frac{1}{S}\sum ^{S}_{s=1}\xi_{c,k}^{MMSE\left (s  \right )}$ and $\overline{\xi}_{k}^{MMSE\left (S  \right )}=\frac{1}{S}\sum ^{S}_{s=1}\xi_{k}^{MMSE\left (s \right )}$, 
where $\xi_{c,k}^{MMSE\left (s  \right )}$ and $\xi_{k}^{MMSE\left (s \right )}$ are associated with the $s$-th realization in $\mathbb{H}^{\left ( S \right )}$. 
The sets of optimum equalizers are defined as $\mathbf{g}_{c,k}^{MMSE}=\big \{ g_{c,k}^{MMSE\left ( s \right )} \mid s \in \mathfrak{S}\big \}$ and $\mathbf{g}_{k}^{MMSE}=\big \{ g_{k}^{MMSE\left ( s \right )} \mid s \in \mathfrak{S}\big \}$.
The sets of optimum weights are  $\mathbf{u}_{c,k}^{MMSE}=\big \{ u_{c,k}^{MMSE\left ( s \right )} \mid s \in \mathfrak{S}\big \}$ and $\mathbf{u}_{k}^{MMSE}=\big \{ u_{k}^{MMSE\left ( s \right )} \mid s \in \mathfrak{S}\big \}$.
From the perspective of each user, the composite optimum equalizers and composite optimum weights are respectively 
\begin{subequations}
\begin{align}
\mathbf{G}^{MMSE} &= \left \{ \mathbf{g}_{c,k}^{MMSE},\mathbf{g}_{k}^{MMSE}\mid k \in \mathcal{K} \right \} \tag{59a}
\\
\mathbf{U}^{MMSE} &= \left \{ \mathbf{u}_{c,k}^{MMSE},\mathbf{u}_{k}^{MMSE}\mid k \in \mathcal{K} \right \}. \tag{59b}
\end{align}
\end{subequations}
Now, we can transform $\mathcal{F}_{RS}^{\left ( S \right )}\left ( P \right )$ into an equivalent WMMSE problem.
\begin{align}
\mathcal{W}_{RS}^{\left ( S \right )}\left ( P \right ): \quad
&
\max_{\overline{\mathbf{c}},\mathbf{P},\mathbf{G},\mathbf{U},\overline{r}_{g},\overline{\mathbf{r}}} \overline{r}_{g}   
\\
s.t. \quad
&\overline{C}_{m}+\overline{r}_{m}\geq \overline{r}_{g},\ \forall m\in\mathcal{M}
\\
&1-\overline{\xi}_{i}^{\left (S  \right )}\geq \overline{r}_{m},\ \forall i \in \mathcal{G}_{m},\  m\in\mathcal{M}
\\
&1-\overline{\xi}_{c,k}^{\left (S  \right )}\geq \sum _{m=1}^{M}\overline{C}_{m},\ \forall k\in\mathcal{K}
\\
&\overline{C}_{m} \geq 0, \ \forall m\in\mathcal{M}
\\
&\mathbf{p}_{c}^{H}\mathbf{D}_{l}\mathbf{p}_{c}+\sum_{m=1}^{M}\mathbf{p}_{m}^{H}\mathbf{D}_{l}\mathbf{p}_{m}\leq P_{l},\ l \in \mathcal{L}
\end{align}
where  $\overline{r}_{g}$ and $\overline{\mathbf{r}}=\left [ \overline{r}_{1} , \cdots,\overline{r}_{M} \right ]$ are auxiliary variables. 
Furthermore, if 
$\big (\mathbf{P}^{*},\mathbf{G}^{*},\mathbf{U}^{*}, \overline{r}_{g}^{*},\overline{\mathbf{r}}^{*},  \overline{\mathbf{c}}^{*} \big )$
satisfies the KKT optimality conditions of
$\mathcal{W}_{RS}^{\left ( S \right )}\left ( P \right )$, 
$\left (\mathbf{P}^{*},  \overline{\mathbf{c}}^{*} \right )$ will satisfy the KKT optimality conditions of $\mathcal{F}_{RS}^{\left ( S \right )}\left ( P \right )$. 
 Note that the WMSEs are convex in each of their corresponding variables (e.g.  equalizers, weights or precoding matrix) when fixing the other two. 
Considering the block-wise convexity property, we use the AO algorithm proposed in \cite{9145200} to solve $\mathcal{W}_{RS}^{\left ( S \right )}\left ( P \right )$. Each iteration is composed of two steps. 
 Variables in the equivalent WMMSE problem are optimized iteratively in an alternating manner. 
The algorithm is guaranteed to converge as the objective function is bounded above for a given power constraint.
 $\overline{r}_{g} $ increases monotonically until convergence as the iteration process goes on. 

\section{Performance Evaluation and Application }
\subsection{Performance Over Rayleigh Channels}


In this section, the performance of RS and NoRS are evaluated over Rayleigh fading channels (representative of conventional cellular terrestrial systems)
with a total transmit power constraint.
During simulation, entries of $\mathbf{H}$ are independently drawn from $\mathcal{CN}\left ( 0,1 \right )$. 
Following the CSIT uncertainty model, entries of $\widetilde{\mathbf{H}}$ are also i.i.d complex Gaussian drawn from $\mathcal{CN}\big ( 0,\sigma _{e}^{2} \big )$, where $\sigma _{e}^{2} =N_{t}^{-1} \sigma _{e,k}^{2} = P^{-\alpha} $. 
Herein, we evaluate the MMF Ergodic Rate by averaging over 100 channel estimates.
For each given channel estimate $\widehat{\mathbf{H}}=\mathbf{H}- \widetilde{\mathbf{H}}$ , its corresponding MMF Average Rate is approximated by SAA method, and the sample size $S$ is set to be 1000. 
$\mathbb{H}^{\left ( S \right )}$ is the set of conditional realizations available at the transmitter. 
The $s$-th conditional realization in $\mathbb{H}^{\left ( S \right )}$ is given by $\mathbf{H}^{\left ( s \right )}=\widehat{\mathbf{H}}+\widetilde{\mathbf{H}}^{\left ( s \right )}$, where $\widetilde{\mathbf{H}}^{\left ( s \right )}$ follows the above CSIT error distribution.


We firstly consider an underloaded system with $N_{t}=6$ transmit antennas, $G=3$ groups and $K=6$ users. The group sizes are respectively $G_{1}=1, \  G_{2}=2, \ G_{3}=3 $. 
Fig. 1 presents the MMF ER of RS and NoRS versus an increasing SNR under various CSIT qualities.
For perfect CSIT, beaming an interference-free stream to each group simultaneously is possible since the system is underloaded. Both RS and NoRS achieve full MMF-DoF and the performance of such two schemes are nearly identical. However, RS shows a little improvement in the rate sense compared with NoRS due to its more flexible architecture. For imperfect CSIT,
the superiority of RS over NoRS becomes more evident.
It can be observed in Fig. 1 that
the MMF-DoF disparity between RS and NoRS gradually appears as the CSIT uncertainty increases. The MMF-DoFs of NoRS and RS in Fig. 1 are respectively $\alpha$ and $\frac{1-\alpha }{M} + \alpha$, which follow the results in Table I. 
This implies that the common stream of RS can provide a DoF gain of $\frac{1-\alpha }{M}$ and consequently MMF rate gains in underloaded regimes,


In Fig. 2, we reduce the number of transmit antennas to 4 and the system becomes partially-overloaded $\left ( K-G_{3}+1\leq N_{t}< K-G_{1}+1  \right )$. 
When considering perfect CSIT, RS and NoRS achieve identical MMF-DoFs at $\frac{1}{2}$. It follows the perfect CSIT results in Table I. Meanwhile, it also follows the results of imperfect CSIT by setting $\alpha = 1$.
Multiplexing gains are partially achieved.
A small rate gap between the two schemes is observed although their MMF-DoFs are equal. 
Next, it comes to imperfect CSIT. We can see that the merit of RS over NoRS becomes more obvious compared with underloaded regime. 
From Fig. 2, the MMF-DoFs of NoRS (blue curves) are approximately $\frac{\alpha }{2}$, which match the theoretical result in (11).
CSIT imperfectness can affect the system performance significantly.
Considering RS, we have $ M_{\mathrm{R}}^{*}= 2$ as a result of $N_{2}\leq N_{t}< N_{3}$ in this specific setup. 
Substituting $M_{\mathrm{R}}^{*}= 2$ and $M = 3$ into (33) or the overloaded results in Table I, we obtain
\begin{equation}
d^{*RS} \geq \left\{
\begin{aligned}
& \frac{1}{2},\quad  0.5 < \alpha \leq 1
\\
& \alpha + \frac{1 - 2\alpha }{3},\quad   0\leq \alpha \leq 0.5.
\end{aligned}
\right.
\end{equation}
In addition, we have $d^{*NoRS} = \frac{\alpha}{2}$.
Such DoF performance is exhibited in Fig. 2. 
All
the simulation results are inline with the theoretical MMF-DoFs in Table I.
Due to the benefits of RS, the system is able to maintain its MMF-DoFs at $\frac{1}{2}$ for all $0.5 < \alpha \leq 1$ in this example.
When $0\leq \alpha \leq 0.5$, the MMF-DoFs decrease slightly to $\alpha + \frac{1 - 2\alpha }{3}$, which is still greater than the $\frac{\alpha }{2}$ achieved by NoRS. 
Compared with the underloaded scenario in Fig. 1, the gaps between RS (red curves) and NoRS (blue curves) increase. 
In other words, the superiority of RS over NoRS becomes more apparent when the system is partially-overloaded.

\begin{figure}
\centering
\begin{minipage}[t]{0.48\textwidth}
\centering
\includegraphics[width=7.3cm]{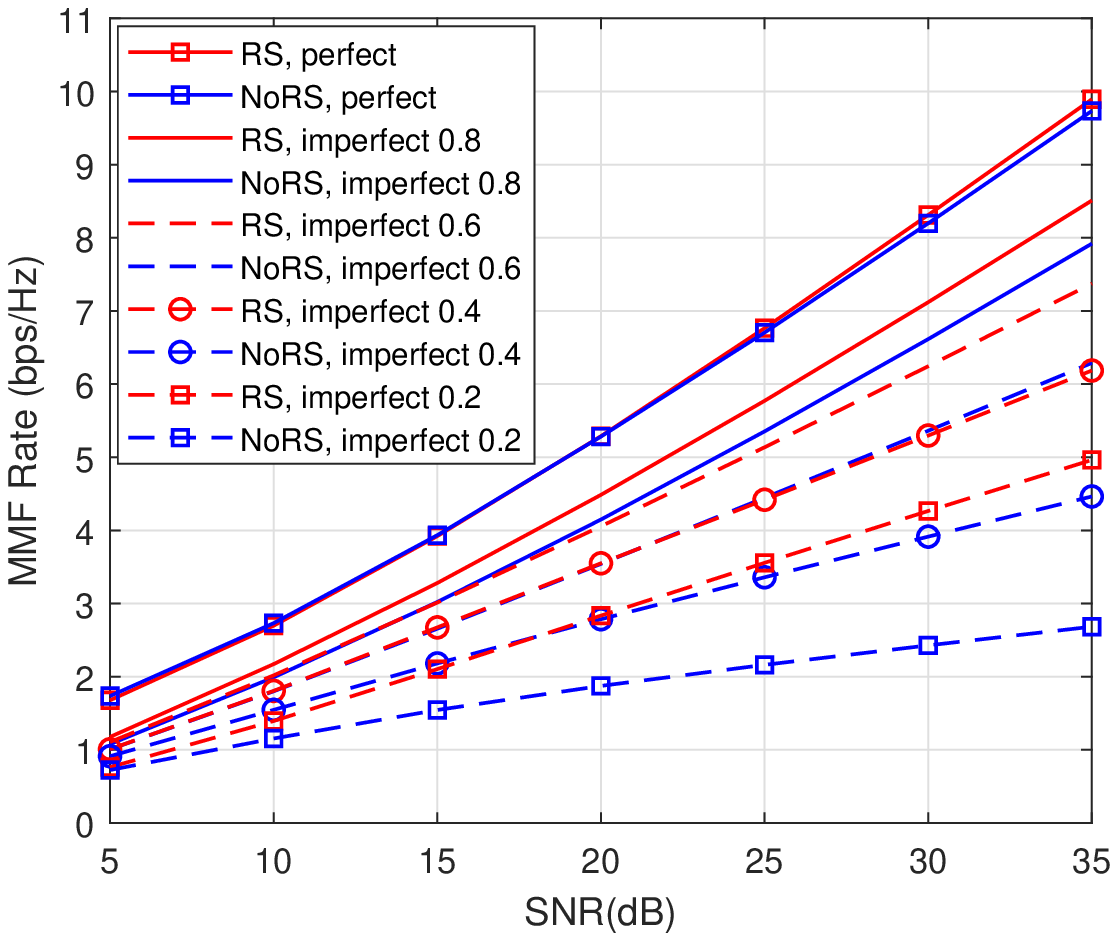}
\caption{MMF rate performance. $N_{t}=6$ antennas, $K=6$ users, $M=3$ groups, $G_{1},G_{2}, G_{3}=1,2,3$ users.}
\end{minipage}
\begin{minipage}[t]{0.48\textwidth}
\centering
\includegraphics[width=7.3cm]{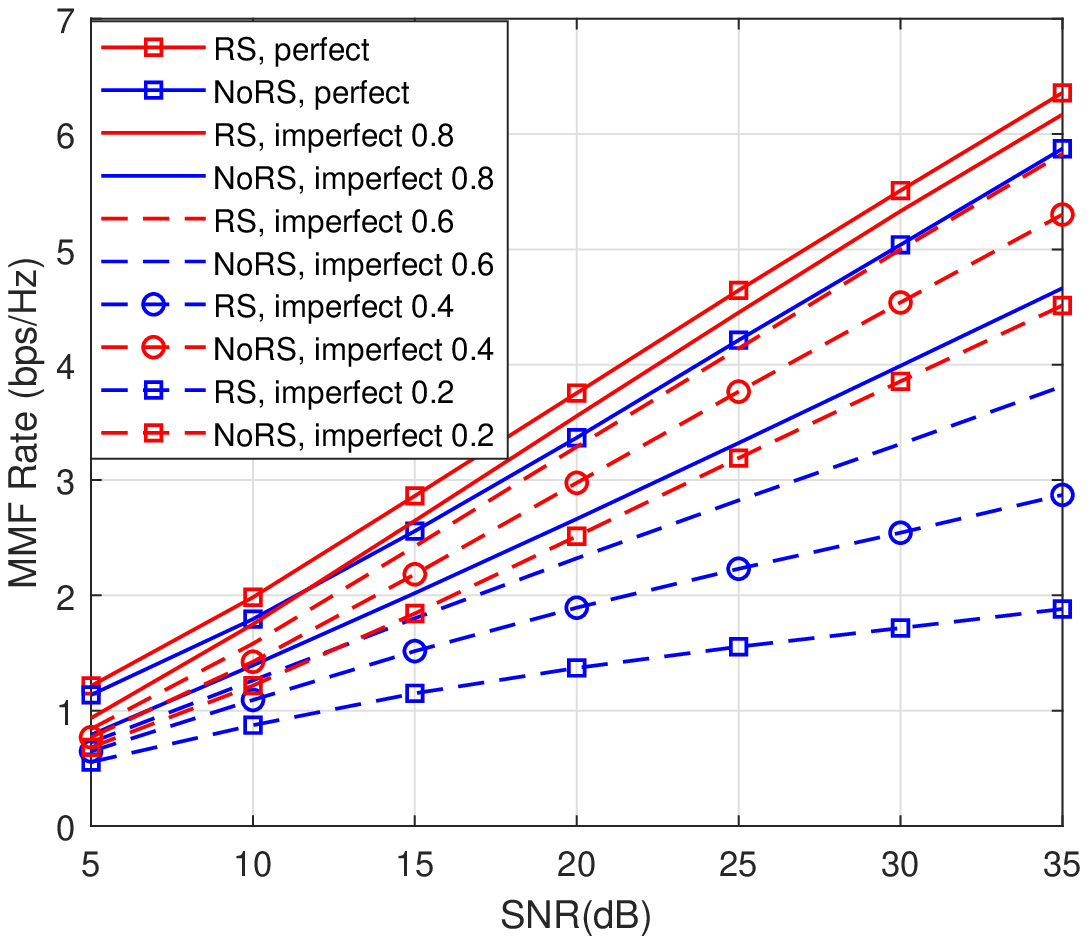}
\caption{MMF rate performance. $N_{t}=4$ antennas, $K=6$ users, $M=3$ groups, $G_{1},G_{2}, G_{3}=1,2,3$ users.}
\end{minipage}
\end{figure}





\begin{figure}
    \centering
    \includegraphics[width=0.42\textwidth]{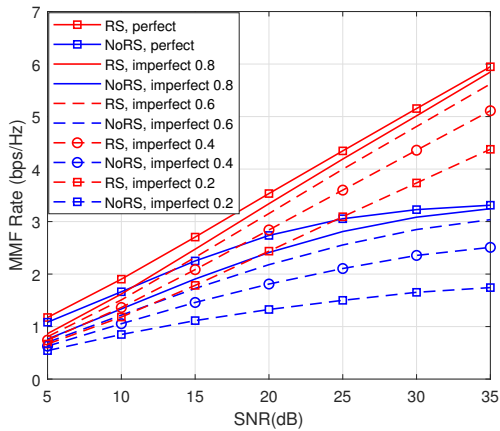}
    \caption{MMF rate performance. $N_{t}=4$ antennas, $K=6$ users, $M=3$ groups, $G_{1},G_{2}, G_{3}=2,2,2$ users.}
    \label{fig:Fig3}
\end{figure}
Furthermore, we keep the same setting as in Fig. 2 but change the group sizes to be symmetric, i.e. $G_{1}=2, \  G_{2}=2, \ G_{3}=2 $. It is noted that the system at present becomes fully-overloaded $\left ( 1 \leq N_{t}< K-G_{3}+1 \right )$.
As illustrated in Fig. 3, RS outperforms NoRS to a great extent in both perfect CSIT and imperfect CSIT scenarios. RS maintains the same MMF-DoFs as in Fig. 2. However, all the multiplexing gains of NoRS are sacrificed and collapse to 0. The corresponding rate performance of NoRS gradually saturates as SNR grows, thus resulting in severe MMF rate limitation. 
Through the simulation results over Rayleigh fading channel, it is demonstrated that RS-based multigroup multicast beamforming is more robust to CSIT imperfectness than the conventional NoRS scheme. 
 RS is able to further exploit spatial multiplexing gains and achieve higher MMF performance in various setups.
 In particular, RS provides significant gains over NoRS in overloaded regimes with imperfect CSIT.
Above all, the gains of RS for multigroup multicast in the presence of imperfect CSIT are shown via simulations in both underloaded and overloaded deployments. This contrasts with \cite{joudeh2017rate} where gains in the presence of perfect CSIT were demonstrated primarily in the overloaded scenarios.

\subsection{Application to Multibeam Satellite Systems}
In order to show the versatility of RS, the application of RS-based multigroup multicast beamforming to multibeam satellite systems is addressed in this section.
Satellite communication, supported by its inherent wide coverage, can not only provide connectivity in unserved areas but also decongest high dense terrestrial networks.
In recent years, the multibeam satellite system
leverages aggressive frequency reuse across multiple narrow spot beams
to support higher throughput \cite{8746876}. 
Note that the framing structure of satellite standard DVB-S2X creates multigroup multicast transmission \cite{DVB}.
Inter-beam interference management techniques need to be implemented.
In this work, 
we focus on a Ka-band multibeam satellite system with multiple single-antenna terrestrial users served by a geostationary orbit (GEO) satellite as shown in Fig. 4.
A single gateway is employed in this system, and the feeder link between gateway and satellite is assumed to be noiseless.
Let $N_{t}$ denote the number of antenna feeds.
The array fed reflector can transform $N_{t}$ feed signals into $M$ transmitted signals (i.e. one signal per beam) to be radiated over the multibeam coverage area \cite{chen2019user}.
Considering single feed per beam (SFPB) architecture which is popular in modern satellites such as Eutelsat Ka-Sat \cite {wang2019multicast, de2017network}, only one feed is required to generate one beam (i.e. $N_{t} = M$). 
Since the multibeam satellite system is in practice user overloaded, we assume that $\rho$ $\left ( \rho > 1 \right )$ users are served simultaneously by each beam. 
Users per beam are uniformly distributed within the satellite coverage area. 
Ideally, the user selection and beamforming can be jointly designed. However this is out of the scope of this paper and can be explored as the future work.
$K = \rho N_{t}$ is the total number of users.

\begin{figure}
    \centering
    \includegraphics[width=0.31\textwidth]{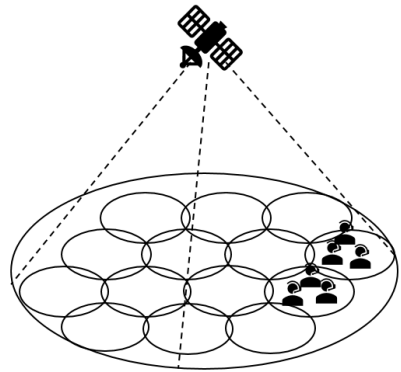}
    \caption{Architecture of multibeam satellite system.}
\end{figure}

\begin{table}
\caption{Multibeam Satellite System Parameter}
\label{table_example}
\centering
\begin{tabular}{c|c}
\hline
\textbf{Parameter} & \textbf{Value}\\
\hline
Frequency band (carrier frequency) & Ka $\left (20\ \mathrm{GHz}  \right )$\\
Satellite height & $35786\ \mathrm{km}\left ( \mathrm{GEO }\right )$\\
User link bandwidth & 500 MHz\\
3 dB angle & $0.4\degree$ \\
Maximum beam gain & 52 dBi\\
User terminal antenna gain & 41.7 dBi\\
System noise temperature &517 K\\
Rain fading parameters & $\left ( \mu,\sigma  \right )=\left ( -3.125,1.591 \right )$\\
\hline
\end{tabular}
\end{table}

\subsubsection{Multibeam Satellite Channel}
The main difference between satellite and terrestrial communications lies in the channel characteristics including free space loss, radiation pattern and atmospheric fading.
Denote $\circ$ as the Hadamard product, and the satellite channel $\mathbf{H} \in \mathbb{C}^{N_{t}\times K}$ is formulated as
\begin{equation}
\mathbf{H} = \mathbf{B} \circ  \mathbf{Q}
\end{equation}
 $\mathbf{B}\in \mathbb{R}^{N_{t}\times K}$ is a matrix composed of receiver antenna gain, free space loss and satellite multibeam antenna gain. Its $\left ( n,k \right )$-th entry can be modeled as 
\begin{equation}
B_{n,k} = \frac{\sqrt{G_{R}G_{n,k}}}{4\pi \frac{d_{k}}{\lambda}\sqrt{\kappa T_{sys}B_{w} }}
\end{equation}
where $G_{R}$ is the user terminal antenna gain, $d_{k}$ is the distance between user-$k$ and the satellite, $\lambda$ is the carrier wavelength, $\kappa$ is the Boltzmann constant, $T_{sys}$ is the receiving system noise temperature and $B_{w}$ denotes the user link bandwidth. 
$G_{n,k}$ is the multibeam antenna gain from the $n$-th feed to the $k$-th user. It mainly depends on the satellite antenna radiation pattern and user locations. In this model, $G_{n,k}$ is approximated by \cite{wang2019multicast}:
\begin{equation}
G_{n,k} = G_{max}\left [\frac{J_{1}\left ( u_{n,k} \right )}{2u_{n,k}} + 36\frac{J_{3}\left ( u_{n,k} \right )}{u_{n,k}^{3}} \right ]^{2}
\end{equation}
where $u_{n,k} = 2.07123\sin \left ( \theta _{n,k} \right )/\sin \left ( \theta _{\mathrm{3dB}} \right )$. Given the $k$-th user position,  $\theta _{n,k}$ is the angle between it and the center of $n$-th beam with respect to the satellite, and $\theta _{\mathrm{3dB}}$ is a 3 dB loss angle compared with the beam center.
The maximum beam gain observed at each beam center is denoted by $G_{max}$.
$J_{1}$ and $J_{3}$ are respectively first-kind Bessel functions with order 1 and order 3. Moreover, the rain fading effect and signal phases are characterized in matrix $\mathbf{\mathbf{Q}}\in \mathbb{C}^{N_{t}\times K}$. Its $\left ( n,k \right )$-th entry is given by
\begin{equation}
Q_{n,k}=   \chi_{k} ^{-\frac{1}{2}} e^{-j\phi_{k}}
\end{equation}
where 
$\chi_{k,dB}=20\log_{10}\left ( \chi_{k} \right )$ is commonly modeled as a lognormal random variable, i.e. $\ln \left (\chi_{k,dB}  \right )\sim  \mathcal{N}\left ( \mu,\sigma   \right )$. 
$\phi_{k}$ is a phase uniformly distributed between 0 and $2\pi$. 
It should be noted that both the fading coefficients and phases are not distinguished among different antenna feeds. 
The reason is that we consider a line-of-sight (LOS) environment and the satellite antenna feed spacing is not large enough compared with the communication distance \cite{wang2019multicast,christopoulos2015multicast,zheng2012generic}. 

\begin{figure}
\centering
\begin{minipage}[t]{0.48\textwidth}
\centering
\includegraphics[width=7.2cm]{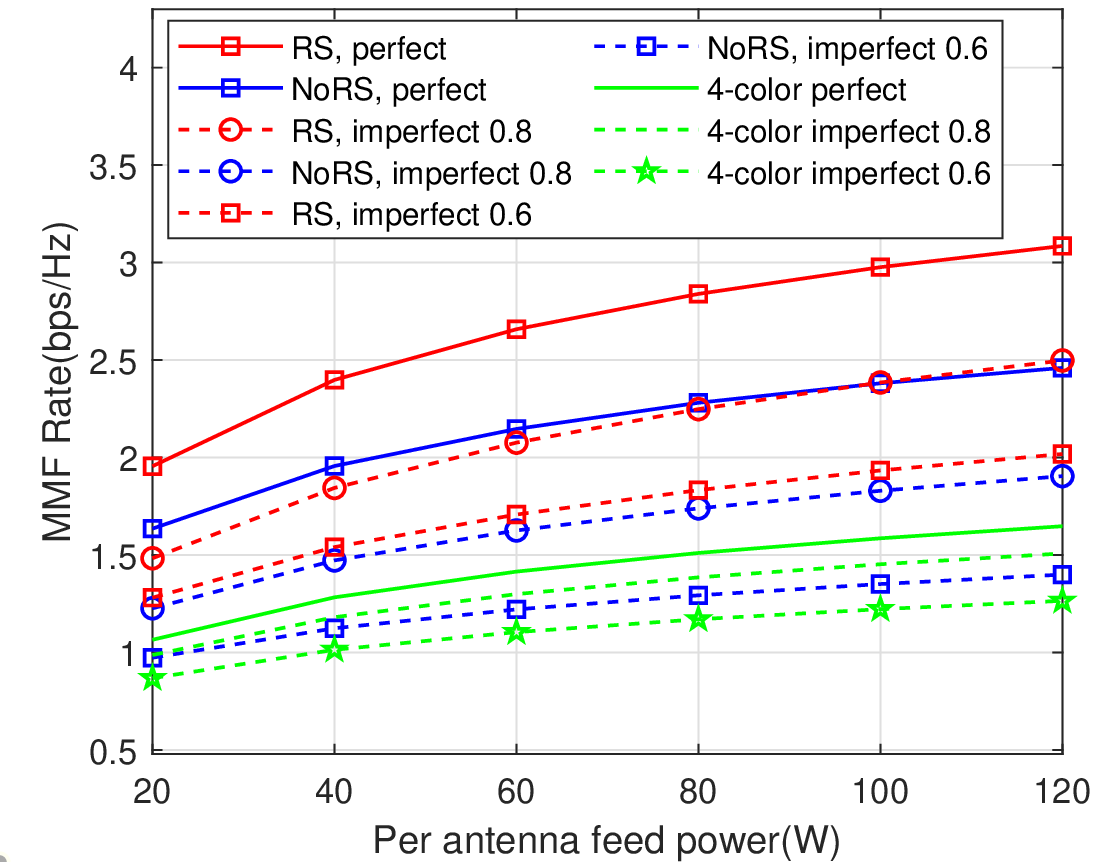}
\caption{MMF rate versus per-feed available power. $N_{t}=7$ antennas, $K=14$ users, $\rho=2$ users.}
\end{minipage}
\begin{minipage}[t]{0.48\textwidth}
\centering
\includegraphics[width=7.2cm]{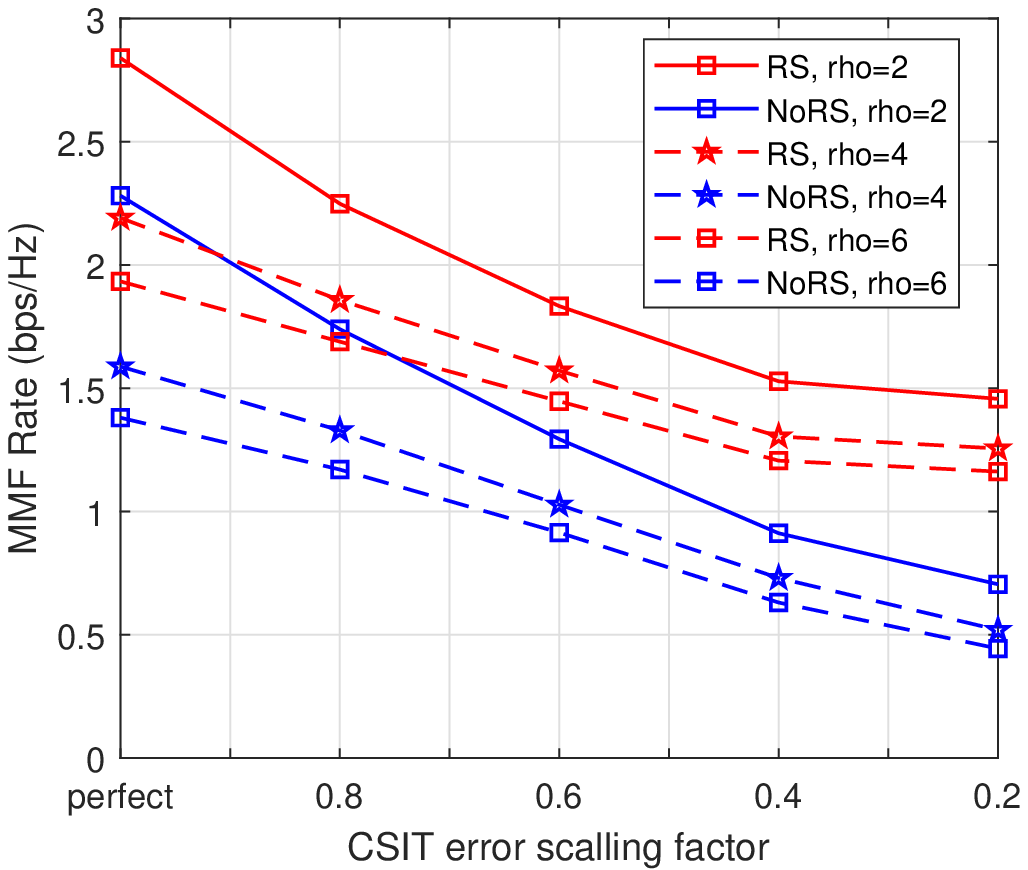}
\caption{MMF rate versus CSIT error scalling factor $\alpha$. $N_{t}=7$ antennas, $\rho=2,\ 4,\ 6$ users, $P/N_{t}=80$ Watts.}
\end{minipage}
\end{figure}

\subsubsection{Performance Over Satellite Channels}
Then, we evaluate the application of RS in multibeam satellite communications.
Results of MMF problems are obtained by averaging 100 satellite channel realizations.
Since non-flexible on-board payloads prevent power sharing between beams, per-feed power constraints are adopted.
System parameters are summarized in Table II.  
Fig. 5 shows the curves of MMF rates among $N_{t}=7$ beams versus an increasing per-feed available transmit power.
We assume two users per beam, i.e. $\rho =2$.
For perfect CSIT, RS achieves around $25\%$ gains over NoRS. For imperfect CSIT, RS is seen to outperform NoRS with $31\%$ and $44\%$ gains respectively when $\alpha = 0.8$ and $\alpha = 0.6$. 
Accordingly, the advantage of employing RS in multigroup multicast beamforming is still observed in multibeam satellite systems.
Through partially
decoding the interference and partially treat the interference
as noise, RS is more robust to the CSIT uncertainty and
overloaded regime than NoRS. Such benefit of RS exactly
tackles the challenges of multibeam satellite communications.
The conventional 4-color scheme performs the worst compared with full frequency reuse schemes.

Fig. 6 depicts the influence of a wider range of CSIT quality on both strategies. 
Here, we set the per-feed available transmit power to be $80\  \mathrm{Watts}$.  As CSIT error scaling factor drops, the MMF rate gap between RS and NoRS increases gradually, which implies the gains of our proposed RS scheme become more and more apparent as the CSIT quality decreases.
In addition, the impact of user number per frame is also studied. 
Since all the users within a beam share the same precoding vector, the beam-rate is determined by the user with the lowest SINR.
Considering $\rho = 2,\ 4,\ 6$ users per frame,
it is clear that  increasing the number of users per frame results in system performance degradation in
both RS and NoRS.

\begin{figure} [htbp]
\centering
\begin{minipage}[t]{0.48\textwidth}
\centering
\includegraphics[width=7.2cm]{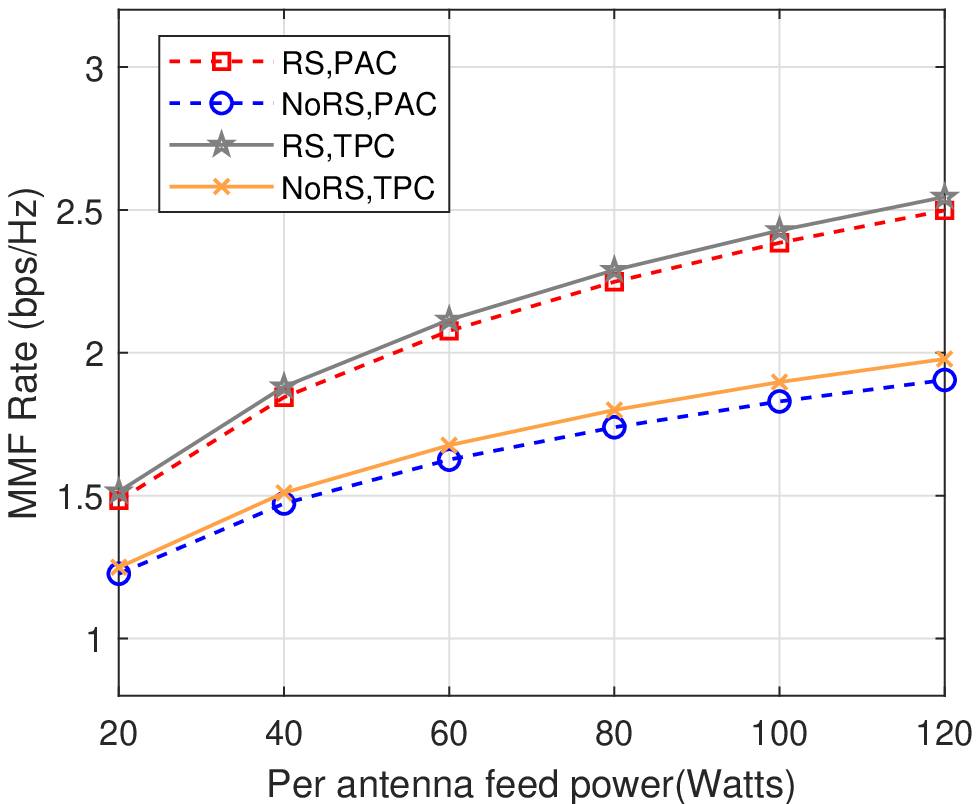}
\caption{MMF rate constrained by PAC/ TPC. $N_{t}=7$ antennas, $K=14$ users, $\rho=2$ users, imperfect CSIT: $\alpha =0.8$.}
\end{minipage}
\begin{minipage}[t]{0.48\textwidth}
\centering
\includegraphics[width=7.2cm]{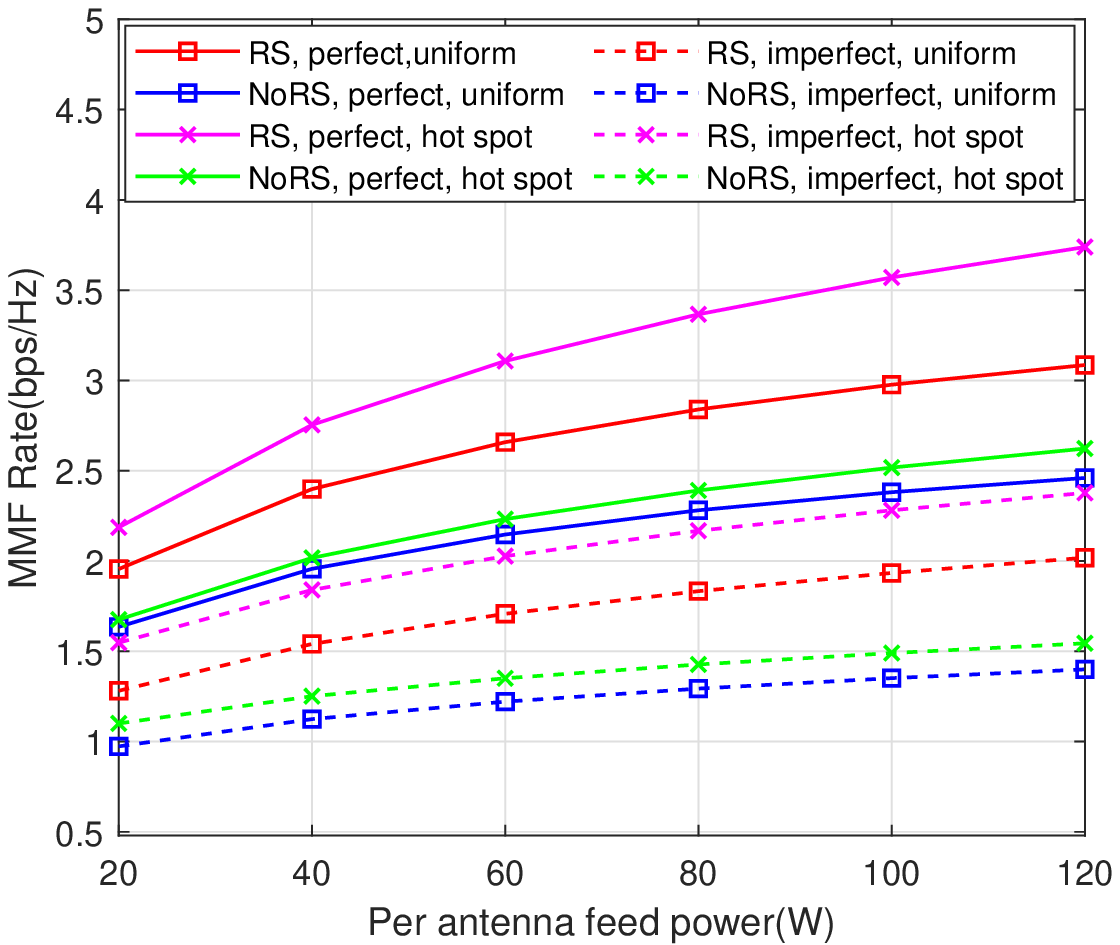}
\caption{MMF rate versus per-feed available power. $N_{t}=7$ antennas, $K=14$ users, imperfect CSIT: $\alpha =0.6$, hot spot $G = \left [ 8,1,1,1,1,1,1 \right ]$.}
\end{minipage}
\end{figure}

Moreover, the impact of different transmit power constraints is studied. Based on the fair PAC assumption, each transmit antenna cannot radiate a power more than $P/N_{t}$. Compared with TPC, the existence of PAC will inevitably restrict the flexibility of beamforming design. 
Taking imperfect CSIT with $\alpha = 0.8$ as an example, Fig. 7 shows the MMF rates when considering TPC and PAC. It is noticed that the practical PAC reduces MMF rate performance slightly in both RS and NoRS.
In Fig. 8, the performance of a hot spot configuration, (e.g. with 8 users in the central beam and 1 user each in the other beams) is compared with the above uniform setting. 
We can observe that the rate improvement provided by RS is more obvious than NoRS, which means that RS is better at managing interference in such hot spot scenario. Specifically, for perfect CSIT, RS outperforms NoRS with $42\%$ gains. For imperfect CSIT, RS achieves higher gains at around $54\%$.




\section{Conclusion}
In this work, we study RSMA for multigroup multicast beamforming in the presence of imperfect CSIT.
Through MMF-DoF analysis, RS is shown to provide gains in both underloaded and overloaded systems compared
with the conventional linear precoding (NoRS).
A generic MMF optimization problem is then formulated and solved by developing a modified WMMSE approach together with an AO algorithm.
The effectiveness of adopting RSMA for multigroup multicast is evaluated through simulations in a wide range of setups. 
Additionally, 
the proposed RSMA framework
is demonstrated very promising for multibeam satellite communications
to manage its inter-beam interference, taking
into account practical challenges such as CSIT uncertainty,
practical per-feed constraints, hotspots, uneven user distribution per beam and overloaded regimes.

\ifCLASSOPTIONcaptionsoff
  \newpage
\fi

\bibliographystyle{IEEEtran}
\bibliography{ref}

\begin{IEEEbiography}[{\includegraphics[width=1in,height=1.25in,clip,keepaspectratio]{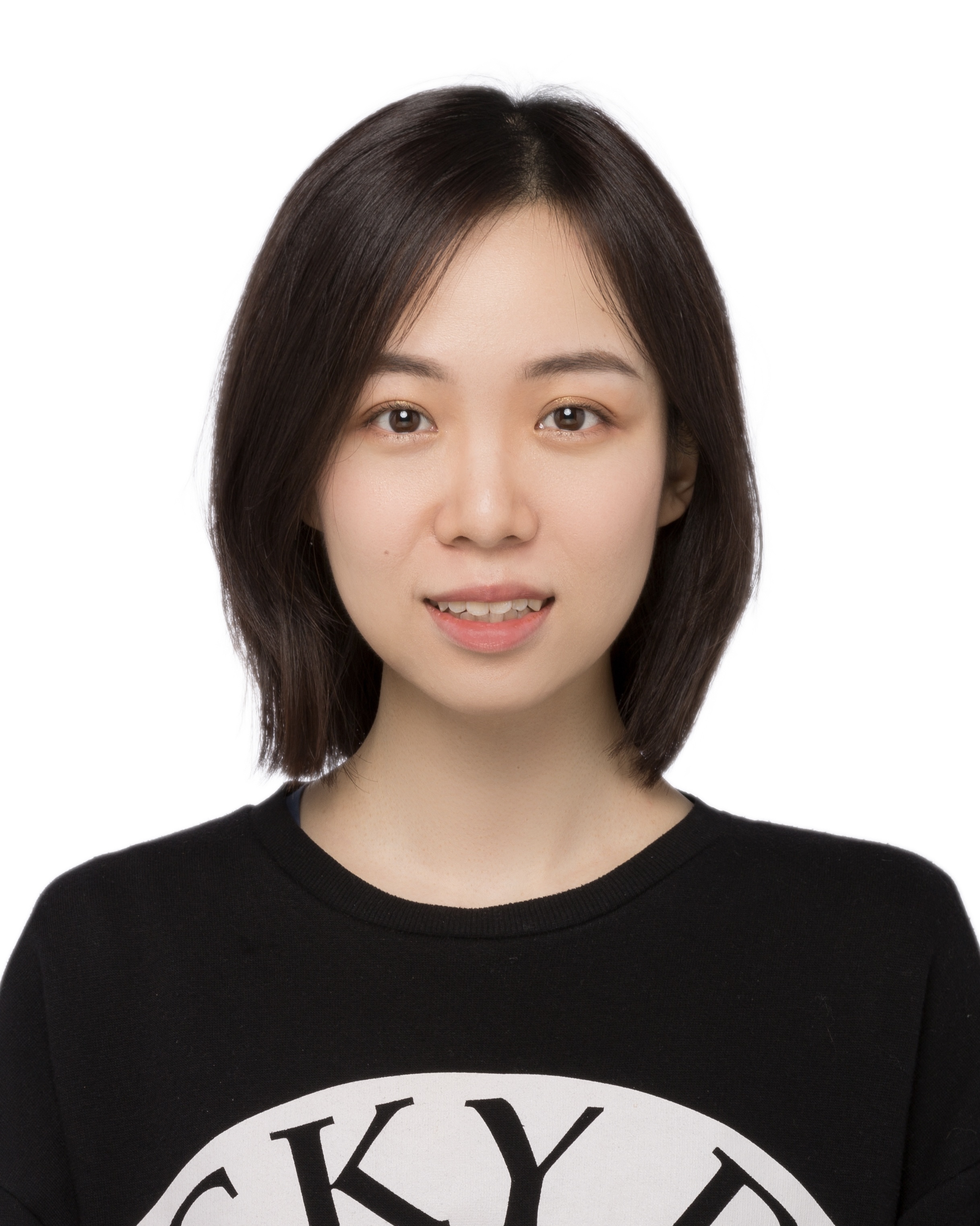}}]{Longfei Yin}
received the B.Eng. degree from Beijing University of Posts and Telecommunications (BUPT), China, in 2017, and the M.Sc. degree in communications and signal processing from Imperial College London, UK, in 2018. 
She is currently a Ph.D. student with the Department of Electrical and Electronic Engineering, Imperial College
London. 
Her research interests include wireless communications, signal processing, rate-splitting multiple access, and satellite communications.
\end{IEEEbiography}

\begin{IEEEbiography}[{\includegraphics[width=1in,height=1.25in,clip,keepaspectratio]{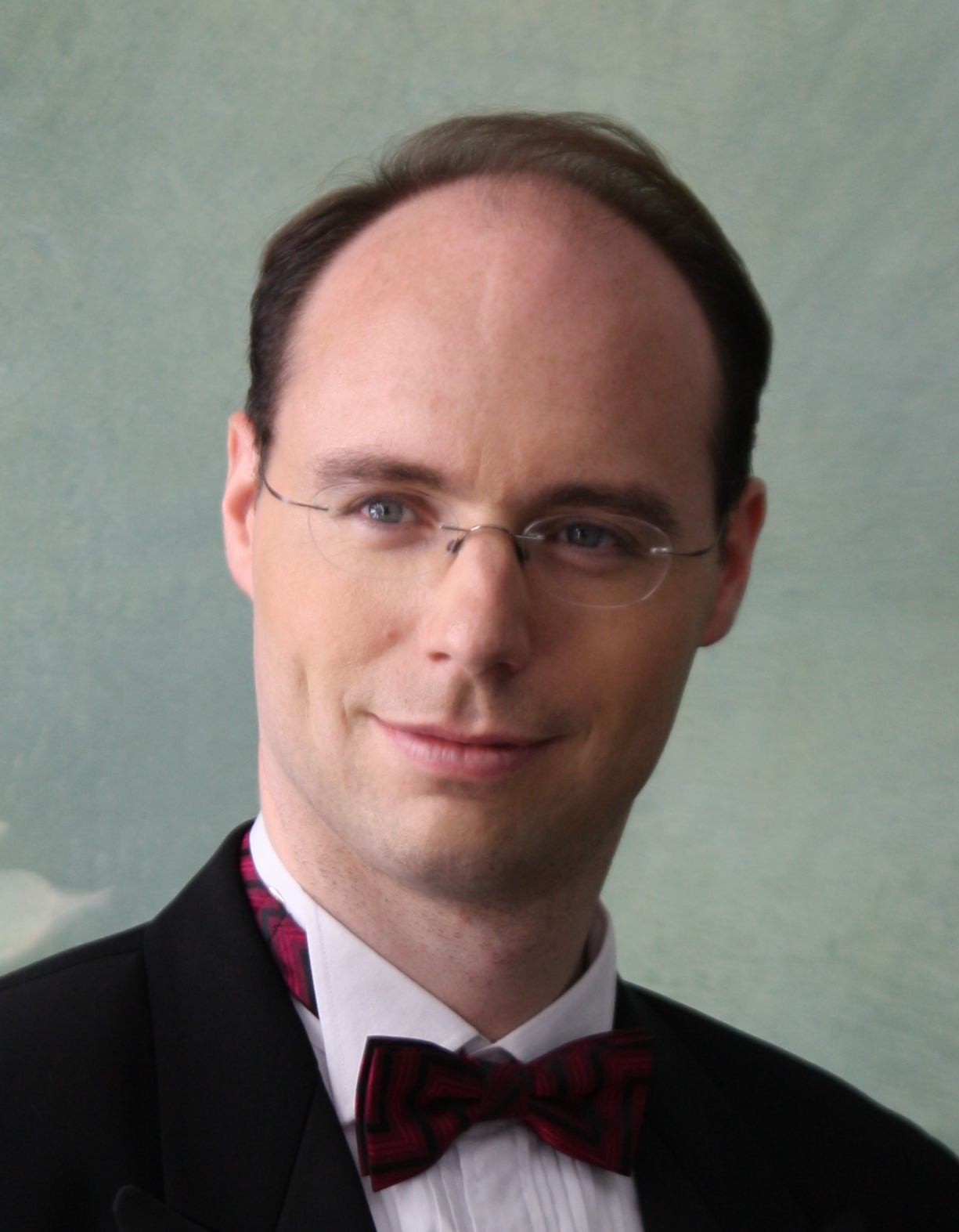}}]{Bruno Clerckx}

is a Professor of Wireless Communications and Signal Processing, the Head of the Wireless Communications and Signal Processing Lab, and the Deputy Head of the Communications and Signal Processing Group, within the Electrical and Electronic Engineering Department, Imperial College London, U.K. He received the M.Sc. and Ph.D. degrees in Electrical Engineering from Université Catholique de Louvain, Belgium, in 2000 and 2005, respectively. From 2006 to 2011, he was with Samsung Electronics, South Korea, where he actively contributed to 4G (3GPP LTE/LTE-A and IEEE 802.16m) and acted as the Rapporteur for the 3GPP Coordinated Multi-Point (CoMP) Study Item. Since 2011, he has been with Imperial College London, first as a Lecturer from 2011 to 2015, Senior Lecturer from 2015 to 2017, Reader from 2017 to 2020, and now as Professor. From 2014 to 2016, he also was an Associate Professor with Korea University, Seoul, South Korea. He also held several visiting positions at Stanford University, EURECOM, National University of Singapore, The University of Hong Kong, Princeton University, The University of Edinburgh, The University of New South Wales, and Tsinghua University.
He has authored two books, 190 peer-reviewed international research papers, and 150 standards contributions, and is the inventor of 80 issued or pending patents among which 15 have been adopted in the specifications of 4G standards and are used by billions of devices worldwide. His research area is communication theory and signal processing for wireless networks. He has been a TPC member, a symposium chair, or a TPC chair of many symposia on communication theory, signal processing for communication and wireless communication for several leading international IEEE conferences. He was an Elected Member of the IEEE Signal Processing Society SPCOM (Signal Processing for Communications and Networking) Technical Committee. He served as an Editor for the IEEE TRANSACTIONS ON COMMUNICATIONS, the IEEE TRANSACTIONS ON WIRELESS COMMUNICATIONS, and the IEEE TRANSACTIONS ON SIGNAL PROCESSING. He has also been a (lead) guest editor for special issues of the EURASIP Journal on Wireless Communications and Networking, IEEE ACCESS, the IEEE JOURNAL ON SELECTED AREAS IN COMMUNICATIONS and the IEEE JOURNAL OF SELECTED TOPICS IN SIGNAL PROCESSING. He was an Editor for the 3GPP LTE-Advanced Standard Technical Report on CoMP.

\end{IEEEbiography}




\end{document}